%% file: main.tex
\documentclass[epj]{svjour}

\input{setup/preamble.tex}

\begin{document}

\input{sections/01_title.tex}
\input{sections/02_abstract.tex}

\maketitle

\input{sections/03_introduction.tex}
\input{sections/04_data.tex}
\input{sections/05_response_definitions.tex}
\input{sections/06_response_implementations.tex}

\input{sections/07_time_shift.tex}
\input{sections/08_short_long.tex}
\input{sections/09_spread_impact.tex}

\input{sections/10_conclusion.tex}

\input{sections/11_paper_contributions.tex}

\begin{appendices}
    \input{sections/12_appendix_A.tex}
\end{appendices}

\bibliographystyle{plain}
\bibliography{bib}

\end{document}

%% file: setup/preamble.tex
\usepackage{latexsym}
\usepackage{graphicx}
\usepackage{hyperref} 
\usepackage[english]{babel}
\usepackage[utf8]{inputenc}
\usepackage{xcolor}
\usepackage{amsmath}
\usepackage{amstext}
\usepackage{url}
\usepackage[english]{babel}
\usepackage{booktabs}
\usepackage{threeparttable}
\usepackage[title]{appendix}
\usepackage{pdflscape}
\usepackage{afterpage}

\hypersetup{%
   	pdfpagelabels=true,%
	plainpages=false,%
	pdfauthor={J. Henao-Londono et al.},%
	pdftitle={Price response functions and spread impact in correlated
			  financial markets},%
	pdfsubject={Complex Systems},%
	bookmarksnumbered=true,%
	colorlinks=true,%
	citecolor=blue,%
	filecolor=black,%
	linkcolor=blue,
	urlcolor=blue,%
	pdfstartview=FitH%
}

%% file: sections/01_title.tex
\title{Price response functions and spread impact in correlated financial
       markets}
\author{Juan C. Henao-Londono
        \thanks{\emph{e-mail: }\href{mailto:juan.henao-londono@uni-due.de}
        {juan.henao-londono@uni-due.de}}
        \and Sebastian M. Krause
        \and Thomas Guhr}
\authorrunning{J. C. Henao-Londono et al.}
\institute{Fakultät für Physik, Universität Duisburg-Essen, Lotharstraße 1,
           47048 Duisburg, Germany}
\date{Received: date / Revised version: date}

%% file: sections/02_abstract.tex
\abstract{
Recent research on the response of stock prices to trading activity revealed
long lasting effects, even across stocks of different companies.
These results imply non-Markovian effects in price formation and when trading
many stocks at the same time, in particular trading costs and price
correlations.
How the price response is measured depends on data set and research focus.
However, it is important to clarify, how the details of the price response
definition modify the results. Here, we evaluate different price response
implementations for the Trades and Quotes (TAQ) data set from the NASDAQ stock
market and find that the results are qualitatively the same for two different
definitions of time scale, but the response can vary by up to a factor of two.
Further, we show the key importance of the order between trade signs and
returns, displaying the changes in the signal strength. Moreover, we confirm
the dominating contribution of immediate price response directly after a trade,
as we find that delayed responses are suppressed. Finally, we test the impact of
the spread in the price response, detecting that large spreads have stronger
impact.
\PACS{
      {89.65.Gh}{Econophysics} \and
      {89.75.-k}{Complex systems} \and
      {05.10.Gg}{Statistical physics}
} 
} 

%% file: sections/03_introduction.tex
\section{Introduction}\label{sec:introduction}

Financial markets use order books to list the number of shares bid or asked at
each price. An order book is an electronic list of buy and sell orders for a
specific security or financial instrument organized by price levels, where
agents can place different types of instructions (orders).

In general, the dynamics of the prices follow a random walk. There are two
extreme models that can describe this behavior: the Efficient Market Hypothesis
(EMH) and the Zero Intelligence Trading (ZIT). The EMH states that all
available information is included in the price and price changes can only be
the result of unanticipated news, which by definition are totally unpredictable
\cite{subtle_nature,Bouchaud_2004,EMH_lillo,stat_theory}. On the other hand,
the ZIT assumes that agents instead of being fully rational, have ``zero
intelligence” and randomly buy or sell. It is supposed that their actions are
interpreted by other agents as potentially containing some information
\cite{subtle_nature,Bouchaud_2004,stat_theory,Wang_2016_cross}.
In both cases the outcome is the same, the prices follow a random walk. Reality
is somewhere in-between \cite{Bouchaud_2004,stat_theory}, and non-Markovian
effects due to stategies or liquidity costs are not contained either.

There are diverse studies focused on the price response
\cite{prop_order_book,dissecting_cross,r_walks_liquidity,subtle_nature,Bouchaud_2004,large_prices_changes,pow_law_dist,theory_market_impact,spread_changes_affect,master_curve,EMH_lillo,quant_stock_price_response,ori_pow_law,Wang_2018_b,Wang_2018_a,Wang_2016_avg,Wang_2016_cross}.
In our opinion, a critical investigation of definitions and methods and how
they affect the results is called for.

Regarding price self-response functions in Refs.
\cite{r_walks_liquidity,subtle_nature,Bouchaud_2004}, Bouchaud et al. found an
increase to a maximum followed by a decrease as the time lag grows.
In Ref. \cite{theory_market_impact}, Gerig found that larger sized transactions
have a larger absolute impact than smaller sized transactions but a much
smaller relative impact. In Ref. \cite{prop_order_book}, it is found that the
impact of small trades on the price is, in relative terms, much larger than
that of large trades and the impact of trading on the price is quasi-permanent.

For price cross-responses functions, Refs.
\cite{dissecting_cross,Wang_2016_cross} revealed that the diagonal terms are on
average larger than the off-diagonal ones by a factor $\sim 5$. The response at
positive times is roughly constant, what is consistent with the hypothesis of a
statistically efficient price. Thus, the current sign does not predict future
returns. In Ref. \cite{Wang_2016_cross} the trends in the cross-responses were
found not to depend on whether or not the stock pairs are in the same economic
sector or extend over two sectors.

Here, we want to discuss, based on a series of detailed empirical results
obtained on trade by trade data, that the variation in the details of the
parameters used in the price response definition modify the characteristics of
the results. Aspects like time scale, time shift, time lag and spread used in
the price response calculation have an influence on the outcomes. To facilitate
the reproduction of our results, the source code for the data analysis is
available in Ref. \cite{code}.

We delve into the key details needed to compute the price response functions,
and explore their corresponding roles. We perform a empirical study in
different time scales. We show that the order between the trade signs and
the returns have a key importance in the price response signal. We split the
time lag to understand the contribution of the immediate returns and the late
returns. Finally, we shed light on the spread impact in the response functions
for single stocks.

The paper is organized as follows: in Sect. \ref{sec:data} we present our data
set of stocks. We then analyze the definition of the price response functions
and describe the physical and trade time in Sect.
\ref{sec:response_functions_def}. We implement different price responses for
several stocks and pairs of stocks in Sect. \ref{sec:response_functions_imp}.
In Sect. \ref{sec:time_shift} we show how the relative position between trade
signs and returns has a huge influence in the results of the computation of the
response functions. In Sect. \ref{sec:short_long} we explain in detail how the
time lag $\tau$ behaves in the response functions. Finally, in Sect.
\ref{sec:spread_impact} we analyze the spread impact in the price response
functions. Our conclusions follow in Sect. \ref{sec:conclusion}.

%% file: sections/04_data.tex
\section{Data set}\label{sec:data}

Modern financial markets, are organized as a double continuous auctions. Agents
can place different types of orders to buy or to sell a given number of shares,
roughly categorized as market orders and limit orders.

In this study, we analyzed trades and quotes (TAQ) data from the NASDAQ stock
market. We selected NASDAQ because it is an electronic exchange where stocks
are traded through an automated network of computers instead of a trading
floor, which makes trading more efficient, fast and accurate. Furthermore,
NASDAQ is the second largest stock exchange based on market capitalization
in the world.

In the TAQ data set, there are two data files for each stock. One gives the
list of all successive quotes. Thus, we have the best bid price, best ask
price, available volume and the time stamp accurate to the second. The other
data file is the list of all successive trades, with the traded price, traded
volume and time stamp accurate to the second. Despite the one second accuracy
of the time stamps, in both files more than one quote or trade may be recorded
in the same second.

To analyze the response functions across different stocks in Sects.
\ref{sec:response_functions_imp}, \ref{sec:time_shift} and
\ref{sec:short_long}, we select the six companies with the largest average
market capitalization (AMC) in three economic sectors of the S\&P index in
2008. Table \ref{tab:companies} shows the companies analyzed with their
corresponding symbol and sector, and three average values for a year.

\begin{table*}[htbp]
\begin{threeparttable}
\caption{Analyzed companies.}
\begin{tabular*}{\textwidth}{c @{\extracolsep{\fill}} ccccc}
\toprule
\bf{Company} & \bf{Symbol} & \bf{Sector} & \bf{Quotes}\tnote{1} &
\bf{Trades}\tnote{2} & \bf{Spread}\tnote{3}\tabularnewline
\midrule
Alphabet Inc. & GOOG & Information Technology (IT) & $164489$ & $19029$ &
$0.40\$$\tabularnewline
Mastercard Inc. & MA & Information Technology (IT) & $98909$ & $6977$ &
$0.38\$$\tabularnewline
CME Group Inc. & CME & Financials (F) & $98188$ & $3032$ &
$1.08\$$\tabularnewline
Goldman Sachs Group Inc. & GS & Financials (F) & $160470$ & $26227$ &
$0.11\$$\tabularnewline
Transocean Ltd. & RIG & Energy (E) & $107092$ & $11641$ &
$0.12\$$\tabularnewline
Apache Corp. & APA & Energy (E) & $103074$ & $8889$ & $0.13\$$\tabularnewline
\bottomrule
\end{tabular*}
\label{tab:companies}
\begin{tablenotes}\footnotesize
\item[1] Average number of quotes from 9:40:00 to 15:50:00 New York time during
 2008.
\item[2] Average number of trades from 9:40:00 to 15:50:00 New York time during
 2008.
\item[3] Average spread from 9:40:00 to 15:50:00 New York time during 2008.
\end{tablenotes}
\end{threeparttable}
\end{table*}

To analyze the spread impact in response functions (Sect.
\ref{sec:spread_impact}), we select 524 stocks in the NASDAQ stock market for
the year 2008. The selected stocks are listed in Appendix
\ref{app:spread_impact}.

In order to avoid overnight effects and any artifact due to the opening and
closing of the market, we systematically discard the first ten and the last
ten minutes of trading in a given day
\cite{Bouchaud_2004,large_prices_changes,spread_changes_affect,Wang_2016_cross}.
Therefore, we only consider trades of the same day from 9:40:00 to 15:50:00
New York local time. We will refer to this interval of time as the ``market
time". The year 2008 corresponds to 253 business days.

%% file: sections/05_response_definitions.tex
\section{Price response function definitions}
\label{sec:response_functions_def}

In Sect. \ref{subsec:key_concepts} we introduce the fundamental quantities used
in the price response definitions. In Sect. \ref{subsec:time_definition} we
describe the physical time scale and the trade time scale. We introduce the
price response functions used in literature in Sect. \ref{subsec:response_def}.

\subsection{Key concepts}\label{subsec:key_concepts}

Market orders are execute at the best available buy or sell price, limit orders
set a maximum purchase price for a buy order, or a minimum sale price for a
sell order. If the limit price is not matched, the order will not be carried
out
\cite{large_prices_changes,predictive_pow,intro_market_micro,stat_theory}.
Limit orders often fail to result in an immediate transaction, and are stored
in a queue called the limit order book
\cite{prop_order_book,stat_prop,predictive_pow,intro_market_micro}. The order
book also identifies the market participants behind the buy and sell orders,
although some choose to remain anonymous. The order book is visible for all
traders and its main purpose is to ensure that all traders have the same
information on what is offered on the market. The order book is the ultimate
microscopic level of description of financial markets.

Buy limit orders are called ``bids", and sell limit orders are called ``asks".
At any given time there is a best (lowest) offer to sell with price
$a\left(t\right)$, and a best (highest) bid to buy with price $b\left(t\right)$
\cite{prop_order_book,subtle_nature,account_spread,limit_ord_spread,stat_theory}.
The price gap between them is called the spread
$s\left(t\right) = a\left(t\right)-b\left(t\right)$
\cite{subtle_nature,market_digest,Bouchaud_2004,account_spread,large_prices_changes,em_stylized_facts,stat_theory}.
Spreads are significantly positively related to price and significantly
negatively related to trading volume. Companies with more liquidity tend to
have lower spreads
\cite{components_spread_tokyo,effects_spread,account_spread,components_spread}.

The average of the best ask and the best bid is the midpoint price, which is
defined as
\cite{prop_order_book,subtle_nature,Bouchaud_2004,large_prices_changes,em_stylized_facts,stat_theory}
\begin{equation}\label{eq:midpoint_price}
    m\left(t\right)=\frac{a\left(t\right)+b\left(t\right)}{2}
\end{equation}
As the midpoint price depends on the quotes, it changes if the quotes change.
The midpoint price grows if the best ask or the best bid grow. On the other
hand, the midpoint price decreases if the best ask or the best bid
decrease.

Price changes are typically characterized as returns. If one denotes
$S\left( t\right)$ the price of an asset at time $t$, the return
$r^{\left(g\right)}\left(t, \tau\right)$, at time $t$ and time lag $\tau$ is
simply the relative variation of the price from $t$ to $t + \tau$
\cite{subtle_nature,empirical_facts,asynchrony_effects_corr,tick_size_impact,causes_epps_effect,non_stationarity},
\begin{equation}\label{eq:return_general}
    r^{\left(g\right)} \left(t, \tau \right) = \frac{S\left(t + \tau\right)
    - S\left(t\right)}{S\left(t\right)}
\end{equation}
It is also common to define the returns as
\cite{dissecting_cross,subtle_nature,empirical_facts,empirical_properties,large_prices_changes,pow_law_dist,theory_market_impact,spread_changes_affect,rand_mat,fluctions_market_friction}
\begin{equation}\label{eq:log_return_general}
    r^{\left(l\right)}\left(t,\tau\right) = \ln S\left(t + \tau\right)
    - \ln S\left(t\right) = \ln \frac{S\left(t + \tau\right)}{S\left(t\right)}
\end{equation}
Equations (\ref{eq:return_general}) and (\ref{eq:log_return_general}) agree if
$\tau$ is small enough \cite{subtle_nature,empirical_facts}.

At longer timescales, midpoint prices and transaction prices rarely differ by
more than half the spread. The midpoint price is more convenient to study
because it avoids problems associated with the tendency of transaction prices
to bounce back and forth between the best bid and ask
\cite{large_prices_changes}.

We define the returns via the midpoint price as
\begin{equation}\label{eq:midpoint_price_return}
    r\left(t,\tau\right) = \frac{m\left(t+\tau\right)-m\left(t\right)}
    {m\left(t\right)}
\end{equation}
The distribution of returns is strongly non-Gaussian and its shape continuously
depends on the return period $\tau$. Small $\tau$ values have fat tails return
distributions \cite{subtle_nature}. The trade signs are defined for general
cases as
\begin{equation}\label{eq:trade_sign_general}
    \varepsilon\left(t\right)=\text{sign}\left(S\left(t\right)
    -m\left(t-\delta\right)\right)
\end{equation}
where $\delta$ is a positive time increment. Hence we have
\begin{equation}\label{eq:trade_sign_results}
    \varepsilon\left(t\right)=\left\{
    \begin{array}{cc}
    +1, & \text{If } S\left(t\right)
    \text{ is higher than the last } m\left( t \right)\\
    -1, & \text{If } S\left(t\right)
    \text{ is lower than the last } m\left( t \right)
    \end{array}\right.
\end{equation}
$\varepsilon(t) = +1$ indicates that the trade was triggered by a market order
to buy and a trade triggered by a market order to sell yields
$\varepsilon(t) = -1$
\cite{subtle_nature,Bouchaud_2004,spread_changes_affect,quant_stock_price_response,order_flow_persistent}.

It is well-known that the series of the trade signs on a given stock exhibit
large autocorrelation. A very plausible explanation of this phenomenon relies
on the execution strategies of some major brokers on given markets. These
brokers have large transactions to execute on the account of some clients. In
order to avoid market making movements because of an inconsiderable large
order, they tend to split large orders into small ones \cite{empirical_facts}.

\subsection{Time definition}\label{subsec:time_definition}

A direct comparison between the trade time scale and the physical time scale is
not possible. To compare them directly we need to assume whether the midpoint
price or the trade signs are on the same scale. We assume the midpoint prices
in the trade time scale to be the same as the midpoint prices in physical time
scale. Therefore, we have the time lag for both computations in seconds. This
approximation allow us to directly compare both scales to have an idea of the
difference and similarities they have. In the other sections, as we are not
directly comparing the time scales, the corresponding quantities of each time
scale are not mixed. Thus physical time scale is measured in seconds and trade
time scale is measured in trades.

Due to the nature of the data, they are several options to define time for
analyzing data.

In general, the time series are indexed in calendar time (hours, minutes,
seconds, milliseconds). Moreover, tick-by-tick data available on financial
markets all over the world is time stamped up to the millisecond, but the order
of magnitude of the guaranteed precision is much larger, usually one second or
a few hundreds of milliseconds \cite{market_digest,empirical_facts}. In several
papers are used different time definitions (calendar time, physical time, event
time, trade time, tick time)
\cite{empirical_facts,sampling_returns,market_making}. The TAQ data used in the
analysis has the characteristic that the trades and quotes can not be directly
related due to the time stamp resolution of only one second
\cite{Wang_2016_cross}. Hence, it is impossible to match each trade with the
directly preceding quote. However, using a classification for the trade signs,
we can compute trade signs in two scales: trade time scale and physical time
scale.

The trade time scale is increased by one unit each time a transaction happens.
The advantage of this count is that limit orders far away in the order book do
not increase the time by one unit. The main outcome of trade time scale is its
``smoothing" of data and the aggregational normality \cite{empirical_facts}.

The physical time scale is increased by one unit each time a second passes.
This means that computing the responses in this scale involves sampling
\cite{sampling_returns,Wang_2016_cross}, which has to be done carefully when
dealing for example with several stocks with different liquidity. This sampling
is made in the trade signs and in the midpoint prices.

Facing the impossibility to relate midpoint prices and trade signs with the TAQ
data in trade time scale, we will use the midpoint price of the previous second
with all the trade signs of the current second. This will be our definition of
trade time scale analysis for the response function analysis.

For physical time scale, as we can sampling, we relate the unique value of
midpoint price of a previous second with the unique trade sign value of the
current second.

\subsubsection*{Trade time scale}\label{subsubsec:trade_time}

We use the trade sign classification in trade time scale proposed in Ref.
\cite{Wang_2016_cross} and used in Refs.
\cite{Wang_2017,Wang_2018_copulas,Wang_2016_avg} that reads

\begin{equation}\label{eq:trade_signs_trade}
    \varepsilon^{\left(t\right)}\left(t,n\right)=\left\{
    \begin{array}{cc}
    \text{sgn}\left(S\left(t,n\right)-S\left(t,n-1\right)\right),
    & \text{if }\\ S\left(t,n\right) \ne S\left(t,n-1\right)\\
    \varepsilon^{\left(t\right)}\left(t,n-1\right),
    & \text{otherwise}
    \end{array}\right.
\end{equation}

$\varepsilon^{\left(t\right)}\left( t,n \right) = +1$ implies a trade triggered
by a market order to buy, and a value
$\varepsilon^{\left(t\right)}\left( t,n \right) = -1$ indicates a trade
triggered by a market order to sell.

In the second case of Eq. (\ref{eq:trade_signs_trade}), if two consecutive
trades with the same trading direction did not exhaust all the available volume
at the best quote, the trades would have the same price, and in consequence
they will have the same trade sign.

With this classification we obtain trade signs for every single trade in the
data set. According to Ref. \cite{Wang_2016_cross}, the average accuracy of the
classification is $85\%$ for the trade time scale.

The TAQ time step is one second, and as it is impossible to find the
correspondences between trades and midpoint prices values inside a second step,
We used the last midpoint price of every second as the representative value of
each second. This introduce an apparent shift between trade signs and returns.
In fact, we set the last midpoint price from the previous second as the first
midpoint price of the current second \cite{Wang_2016_cross}.

\begin{figure}[htbp]
    \centering
    \includegraphics[width=\columnwidth]
    {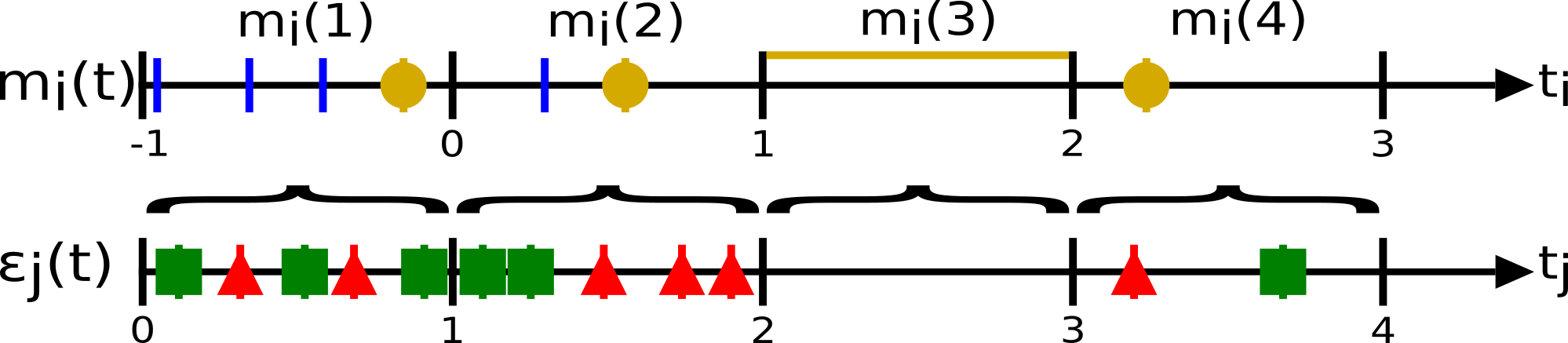}
    \caption{Sketch of data processing for trade time scale. In the midpoint
             price time line, the vertical lines represent the change in price
             of the quotes and the circles represent the last price change in a
             quote in a second. In the trade signs time line, the squares
             represent the buy market orders and the triangles represent the
             sell market orders. The midpoint price time line and the trade
             sign time line are shifted in one second.}
    \label{fig:relation_trades_midpoint_trade_scale}
\end{figure}

As we know the second in which the trades were made, we can relate the trade
signs and the midpoint prices as shown in Fig.
\ref{fig:relation_trades_midpoint_trade_scale}. For the trade time scale, there
are in general, several midpoint prices in a second. For each second we select
the last midpoint price value, and we relate it to the next second trades. In
Fig. \ref{fig:relation_trades_midpoint_trade_scale}, the last midpoint price
(circle) between the second $-1$ and $0$ is related to all the trades (squares
and triangles) in the second $0$ to $1$, and so on. In the seconds when the
quotes do not change, the value of the previous second (vertical line over the
physical time interval) is used. Thus, all the seconds in the open market time
have a midpoint price value, and in consequence returns values. We assume that
as long as no changes occurred in the quotes, the midpoint price remains the
same as in the previous second.

The methodology described is an approximation to compute the response in the
trade time scale. A drawback in the computation could come from the fact that
the return of a given second is composed by the contribution of small returns
corresponding to each change in the midpoint price during a second. As we are
assuming only one value for the returns in each second, we consider all the
returns in one second interval to be positive or negative with the same
magnitude, which could not be the case. This could increase or decrease the
response signal at the end of the computation.

\begin{figure}[htbp]
    \centering
    \includegraphics[width=\columnwidth]{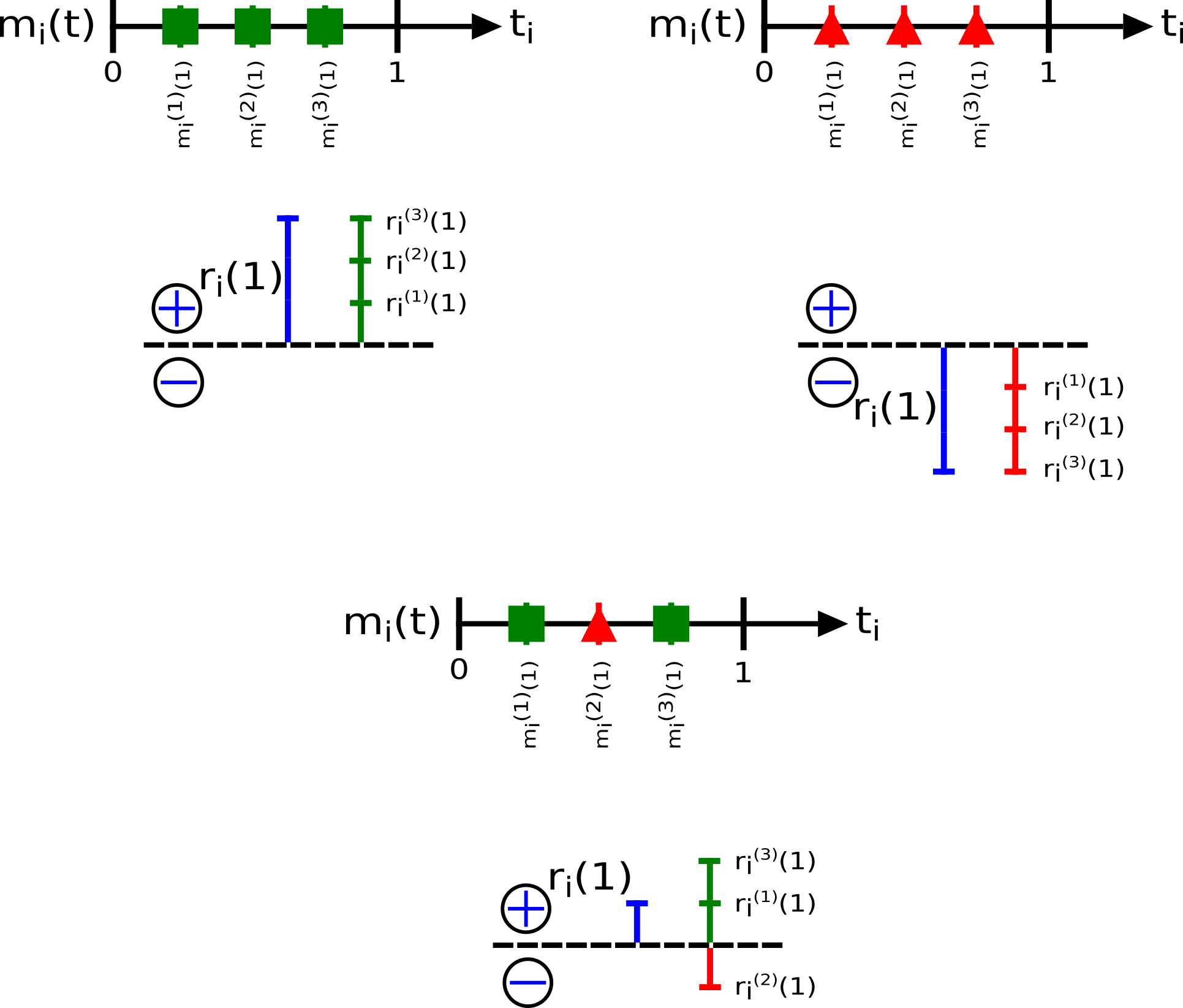}
    \caption{Sketch of the return contributions from every midpoint price
             change in a second. The squares represent the rise of the price of
             the midpoint price and the triangles represent the decrease of the
             price of the midpoint price. We illustrate three cases: (top left)
             the changes of the midpoint prices and return are due to the rise
             of the prices, (top right) the changes of the midpoint prices and
             return are due to the decrease of the prices, and (bottom) the
             changes of the midpoint prices and return are due to a combination
             of rise and decrease of the prices. The blue vertical line
             represents the net return in each case.}
    \label{fig:return_contributions}
\end{figure}

Figure \ref{fig:return_contributions} illustrate this point. Suppose there are
three different midpoint prices in one second interval and thus, three
different returns for these three midpoint price values. Furthermore, suppose
that the volume of limit orders with the corresponding midpoint prices are the
same in the bid and in the ask (the returns have the same magnitude). In the
case of the top left (top right) sketch, all the changes are due to the rise
(decrease) of the midpoint price, that means, consumption of the best ask
(bid), so all the contributions of the individual returns in the second are
positive (negative), and in consequence, the net return is positive (negative).
In the case of the bottom, the changes are due to a combination of increase and
decrease of the midpoint price, so in the end, the individual returns sum up to
a net return, which can be positive or negative, depending of the type of
midpoint price values in the interval. Thus, in this case, we are assuming in
the end that all the returns were positive or negative, which probably was not
the case, and in consequence will increase or decrease the real value of the
net return.

In all cases, we choose the last change in the midpoint price in a second
interval as described before in Fig.
\ref{fig:relation_trades_midpoint_trade_scale}. We use this method knowing that
the variation in one second of the midpoint price is not large (in average, the
last midpoint price of a second differ with the average midpoint of that second
in $0.007\%$), hence it can give us representative information on the response
functions.

\subsubsection*{Physical time scale}\label{subsubsec:physical_time}

We use the trade sign definition in physical time scale proposed in Ref.
\cite{Wang_2016_cross} and used in Refs.
\cite{Wang_2017,Wang_2016_avg}, that depends on the classification in
Eq. (\ref{eq:trade_signs_trade}) and reads
\begin{equation}\label{eq:trade_signs_physical}
    \varepsilon^{\left(p\right)}\left(t\right)=\left\{
    \begin{array}{cc}
    \text{sgn}\left(\sum_{n=1}^{N\left(t\right)}\varepsilon^{\left(t\right)}
    \left(t,n\right)\right),
    & \text{If }N \left(t\right)>0\\
    0, & \text{If }N\left(t\right)=0
    \end{array}\right.
\end{equation}
where $N \left(t \right)$ is the number of trades in a second interval.
$\varepsilon^{\left(p\right)}\left( t \right) = +1$ implies that the majority
of trades in second $t$ were triggered by a market order to buy, and a value
$\varepsilon^{\left(p\right)}\left( t \right) = -1$ indicates a majority of
sell market orders. In this definition, they are two ways to obtain
$\varepsilon^{\left(p\right)}\left( t \right) = 0$. One way is that in a
particular second there are no trades, and then no trade sign. The other way is
that the addition of the trade signs ($+1$ and $-1$) in a second be equal to
zero. In this case, there is a balance of buy and sell market orders.

Market orders show opposite trade directions to limit order executed
simultaneously. An executed sell limit order corresponds to a buyer-initiated
market order. An executed buy limit order corresponds to a seller-initiated
market order.

As the trade time scale, on the physical time scale we use the same strategy
to obtain the midpoint price for every second, so all the seconds in the open
market time have a midpoint price value. Even if there is no change of quotes
in a second, it still has a midpoint price value and return value.

\begin{figure}[htbp]
    \centering
    \includegraphics[width=\columnwidth]
    {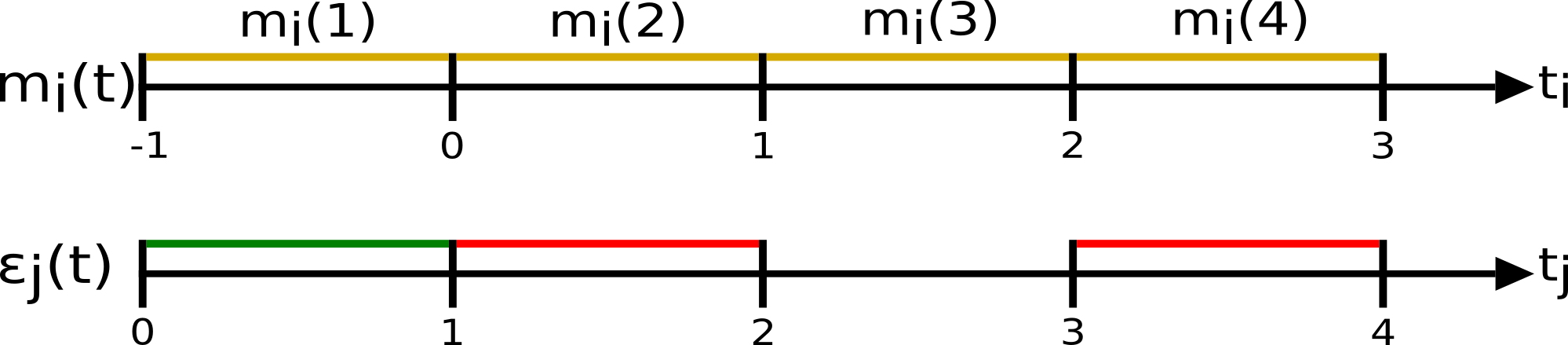}
    \caption{Sketch of data processing for physical time scale. In the midpoint
             price time line, the horizontal lines between seconds represent
             the midpoint prices. In the trade signs time line, the horizontal
             lines between seconds represent the trade sign values. The
             midpoint price time line and the trade sign time line are shifted
             in one second.}
    \label{fig:relation_trades_midpoint_time_scale}
\end{figure}

In this case we do not compare every single trade sign in a second, but the net
trade sign obtained for every second with the definition, see
Eq. (\ref{eq:trade_signs_physical}). This can be seen in
Fig. \ref{fig:relation_trades_midpoint_time_scale}, where we related the
midpoint price of the previous second with the trade sign of the current
second. According to Ref. \cite{Wang_2016_cross}, this definition has an
average accuracy up to $82\%$ in the physical time scale.


\subsection{Response function definitions}\label{subsec:response_def}

The response function measures price changes resulting from execution of market
orders. In Refs.
\cite{r_walks_liquidity,subtle_nature,Bouchaud_2004}, Bouchaud et al. use a
self-response function that only depends on the time lag $\tau$. This function
measures how much, on average, the price moves up (down) at time $\tau$
conditioned to a buy (sell) order at time zero. They found for France Telecom
that the response function increases by a factor $~2$ between $\tau = 1$ and
$\tau \approx 1000$ trades, before decreasing back. For larger $\tau$, the
response function decreases, and even becomes negative beyond
$\tau \approx 5000$. However, in some cases the maximum is not observed and
rather the price response function keeps increasing mildly
\cite{subtle_nature}.

In Ref. \cite{theory_market_impact}, the price impact function, is defined as
the average price response due to a transaction as a function of the
transaction's volume. Empirically the function is highly concave
\cite{theory_market_impact}. The curvature of the price impact function is
entirely due to the probability that a transaction causes a nonzero impact. The
larger the size of the transaction, the larger the probability. In Ref.
\cite{prop_order_book}, they found that the response function for three French
stocks first increases from $\tau = 10s$ to a few hundred seconds, and then
appears to decrease back to a finite value.

In Ref. \cite{dissecting_cross} they defined a response function who measures
the average price change of a contract $i$ at time $t+\tau$, after experiencing
a sign imbalance in contract $j$ at time $t$. In this work $\tau$ is used in
units of five minutes.

In later works \cite{spread_changes_affect,Wang_2016_cross}, Grimm et al.
and Wang et al. use the logarithmic return for stock $i$ and time lag $\tau$,
defined via the midpoint price $m_{i} \left( t \right)$ to define a
cross-response function. The response function measures how a buy or sell order
at time $t$ influences on average the price at a later time $t + \tau$.
The physical time scale was chosen since the trades in different stocks are not
synchronous (TAQ data). They found that in all cases, an increase  to a
maximum is followed by a decrease. The trend is eventually reversed.

Finally, in Ref. \cite{Wang_2018_b}, Wang et al. define the response function
on a trade time scale (Totalview data), as the interest is to analyze the
immediate responses. Here, the time lag $\tau$ is restricted to one trade, such
that the price response quantifies the price impact of a single trade.

%% file: sections/06_response_implementations.tex
\section{Price response function implementations}
\label{sec:response_functions_imp}

The main objective of this work is to analyze the price response functions. In
general we define the self- and cross-response functions in a correlated
financial market as
\begin{equation}\label{eq:response_general}
    R^{\left(scale\right)}_{ij}\left(\tau\right)=\left\langle
    r^{\left(scale\right)}_{i}\left(t-1, \tau\right)
    \varepsilon^{\left(scale\right)}_{j} \left(t\right)\right\rangle_{average}
\end{equation}
where the index $i$ and $j$ correspond to stocks in the market,
$r^{\left(scale\right)}_{i}$ is the return of the stock $i$ in a time lag
$\tau$ in the corresponding scale and $\varepsilon^{\left(scale\right)}_{j}$ is
the trade sign of the stock $j$ in the corresponding scale. The superscript
$scale$ refers to the time scale used, whether physical time scale
($scale = p$) or trade time scale ($scale = t$). Finally, The subscript
$average$ refers to the way to average the price response, whether relative to
the physical time scale ($average = P$) or relative to the trade time scale
($average = T$).

We use the returns and the trade signs to define three response functions:
trade time scale response, physical time scale response and activity response.

To compare the three response functions, we define the following quantities
\begin{align}
    E_{j,d}\left(t\right)&=\sum_{n=1}^{N\left(t\right)}
    \varepsilon_{j,d}^{\left(t\right)}\left(t,n\right) =
    \text{sgn}\left(E_{j,d}\left(t\right)\right)
    \left|E_{j,d}\left(t\right)\right|\label{eq:sum_trades}\\
    \varepsilon_{j,d}^{\left(p\right)}\left(t\right)&=
    \text{sgn}\left(E_{j,d}\left(t\right)\right)\label{eq:sign_sum_trades}
\end{align}
where the subscript $d$ refers to the days used in the response computation.
We use Eq. (\ref{eq:sum_trades}) to make easier the comparison between the
results, as all the defined responses use the trade sign term.

In Sect. \ref{subsec:response_function_trade} we analyze the responses
functions in trade time scale, in Sect. \ref{subsec:response_function_physical}
we analyze the responses functions in physical time scale and in Sect.
\ref{subsec:activity_response_function} we define a response function to
analyze the influence of the frequency of trades in a second.

\subsection{Response functions on trade time scale}
\label{subsec:response_function_trade}

We define the self- and cross-response functions in trade time scale, using the
trade signs in trade time scale. For the returns, we select the last midpoint
price of every second and compute them. We use this strategy with the TAQ data
set considering that the price response in trade time scale can not be directly
compared with the price response in physical time scale. In this case we relate
each trade sign in one second with the midpoint price of the previous second.
Then, to compute the returns, instead of using trades as the time lag (it would
make no sense as all the midpoint price are the same in one second) we use
seconds. Thus, we force the response in trade time scale to have a physical
time lag, and then, be able to compare with the physical time scale response.
This approximation is feasible considering the discussion in Sect.
\ref{sec:response_functions_def}. The price response function in trade time
scale is defined as
\begin{equation}\label{eq:response_functions_trade_scale_general}
    R^{\left(t\right)}_{ij}\left(\tau\right)=\left\langle r^{\left(p\right)}
    _{i}\left(t-1,\tau \right)\varepsilon_{j}^{\left(t\right)}
    \left(t, n\right)\right\rangle _{T}
\end{equation}
where the superscript $t$ refers to the trade time scale. We explicitly
calculate the average in Eq. (\ref{eq:response_functions_trade_scale_general}),
\begin{align}\label{eq:response_trades_explicit}
    R_{ij}^{\left(t\right)}\left(\tau\right)&=\frac{1}{\sum_{d=d_{0}}^{d_{f}}
    \sum_{t=t_{0}}^{t_{f}}N_{j,d} \left(t \right)} \nonumber \\
    &\sum_{d=d_{0}}^{d_{f}}\sum_{t=t_{0}}^{t_{f}}\sum_{n=1}
    ^{N\left(t\right)} r^{\left(p\right)}_{i,d}\left(t-1, \tau\right)
    \varepsilon_{j,d}^{\left(t\right)}\left(t,n\right)\\
    &=\sum_{d=d_{0}}^{d_{f}}\sum_{t=t_{0}}^{t_{f}} r^{\left(p\right)}_{i,d}
    \left(t-1,\tau\right) \frac{\sum_{n=1}^{N\left(t\right)}
    \varepsilon_{j,d}^{\left(t\right)}\left(t,n \right)}
    {\sum_{d=d_{0}}^{d_{f}} \sum_{t=t_{0}}^{t_{f}}N_{j,d}\left(t\right)}
    \nonumber \\
    &=\sum_{d=d_{0}}^{d_{f}}\sum_{t=t_{0}}^{t_{f}}r^{\left(p\right)}_{i,d}
    \left(t-1,\tau\right) \text{sgn}\left(E_{j,d}\left(t\right)\right)
    w_{j,d}^{\left(t\right)}\left(t\right)
\end{align}
where
\begin{equation}\label{eq:trade_weight}
    w_{j,d}^{\left(t\right)}\left(t\right) =
    \frac{\left|E_{j,d}\left(t\right)\right|}{\sum_{d=d_{0}}^{d_{f}}
    \sum_{t=t_{0}}^{t_{f}}N_{j,d} \left(t\right)}
\end{equation}
is a weight function that depends on the normalization of the response.

To compute the response functions on trade time scale, we used all the trade
signs during a day in market time. As we can not associate an individual
midpoint price with their corresponding trade signs, all the trade signs in one
second are associated with the midpoint price of the previous second.
As $\tau$ depends on the midpoint price, even if we are using trade signs in
trade time scale, the value of $\tau$ is in seconds.

\begin{figure*}[htbp]
    \centering
    \includegraphics[width=\textwidth]
    {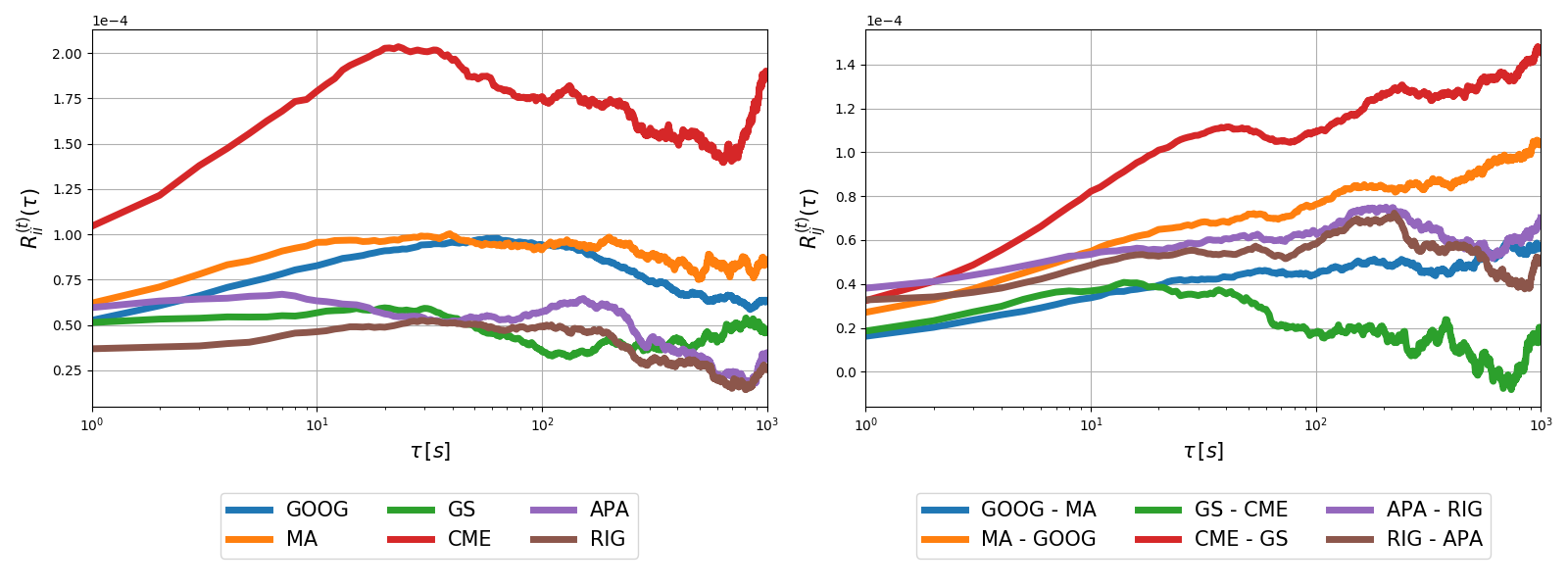}
    \caption{Self- and cross-response functions
             $R^{\left(t\right)}_{ij}\left(\tau\right)$ in 2008 versus time lag
             $\tau$ on a logarithmic scale in trade time scale. Self-response
             functions (left) of individual stocks and cross-response functions
             (right) of stock pairs from the same economic sector.}
    \label{fig:response_function_trade_scale}
\end{figure*}

The results of Fig. \ref{fig:response_function_trade_scale} show the self-
responses of the six stocks used in the analysis and the cross-responses for
pairs of stocks representing three different economic sectors.

The self-response functions increase to a maximum and then slowly decrease. In
some stocks this behavior is more pronounced than in others. For our selected
tickers, a time lag of $\tau = 10^{3}s$ is enough to see an increase to a
maximum followed by a decrease. Thus, the trend in the self-response functions
is eventually reversed.
On the other hand, the cross-response functions have smaller signal strength
than the self-response functions. For our cross-response functions of stocks in
the same sectors, some couples exhibit the increase-decrease behavior inside a
time lag of $\tau = 10^{3}s$. Other couples seems to need a larger time lag to
reach the decrease behavior.

\subsection{Response functions on physical time scale}
\label{subsec:response_function_physical}

One important detail to compute the market response in physical time scale is
to define how the averaging of the function will be made, because the
response functions highly differ when we include or exclude
$\varepsilon^{\left(p\right)}_j \left( t\right) = 0$ \cite{Wang_2016_cross}.
The price responses including
$\varepsilon^{\left(p\right)}_j \left( t\right) = 0$ are weaker than the
excluding ones due to the omission of direct influence of the lack of trades.
However, either including or excluding $\varepsilon^{p}_j \left( t\right) = 0$
does not change the trend of price reversion versus the time lag, but it does
affect the response function strength \cite{Wang_2016_avg}. For a deeper
analysis of the influence of the term
$\varepsilon^{\left(p\right)}_j \left( t\right) = 0$, we suggest to check Refs.
\cite{Wang_2016_avg,Wang_2016_cross}. We will only take in account the
response functions excluding $\varepsilon^{p}_j \left( t\right) = 0$.

We define the self- and cross-response functions in physical time scale, using
the trade signs and the returns in physical time scale. The price response
function on physical time scale is defined as
\begin{equation}\label{eq:response_functions_time_scale_general}
    R^{\left(p\right)}_{ij}\left(\tau\right)=\left\langle r^{\left(p\right)}
    _{i}\left(t-1, \tau\right) \varepsilon_{j}^{\left(p\right)} \left(t\right)
    \right\rangle _{P}
\end{equation}
where the superscript $p$ refers to the physical time scale. The explicit
expression corresponding to Eq.
(\ref{eq:response_functions_time_scale_general}) reads
\begin{align}\label{eq:response_seconds_explicit}
    R_{ij}^{\left(p\right)}\left(\tau\right)&=\frac{1}{\sum_{d=d_{0}}^{d_{f}}
    \sum_{t=t_{0}}^{t_{f}} \eta\left[ \varepsilon_{j,d}^{\left(p\right)}
    \left(t\right)\right]} \nonumber \\
    &\sum_{d=d_{0}}^{d_{f}} \sum_{t=t_{0}}^{t_{f}}
    r^{\left(p\right)}_{i,d}\left(t-1,\tau\right)
    \varepsilon_{j,d}^{\left(p\right)}\left(t\right)
    \eta\left[\varepsilon_{j}^{\left(p\right)} \left(t\right)\right] \\
    &=\sum_{d=d_{0}}^{d_{f}}\sum_{t=t_{0}}^{t_{f}}r^{\left(p\right)}_{i,d}
    \left(t-1,\tau\right) \frac{\varepsilon_{j,d}^{\left(p\right)}
    \left(t\right) \eta\left[\varepsilon_{j,d}^{\left(p\right)}
    \left( t\right)\right]} {\sum_{d=d_{0}}^{d_{f}}\sum_{t=t_{0}}^{t_{f}}\eta
    \left[\varepsilon_{j,d}^{\left(p\right)} \left(t\right)\right]} \nonumber\\
    &=\sum_{d=d_{0}}^{d_{f}}\sum_{t=t_{0}}^{t_{f}}r^{\left(p\right)}_{i,d}
    \left(t-1,\tau\right) \text{sgn}\left(E_{j,d}\left(t\right)\right)
    w_{j,d}^{\left(p\right)}\left(t\right)
\end{align}
where
\begin{equation}
    \eta\left(x\right)=\left\{ \begin{array}{cc}
    1, & \text{If }x\ne0 \\
    0, & \text{otherwise}
    \end{array}\right.
\end{equation}

take only in account the seconds with trades and
\begin{equation}
    w_{j,d}^{\left(p\right)}\left(t\right) = \frac{\eta\left[\text{sgn}
    \left(E_{j,d}\left( t\right)\right)\right]}{\sum_{d=d_{0}}^{d_{f}}
    \sum_{t=t_{0}}^{t_{f}} \eta\left[\text{sgn}\left(E_{j,d}
    \left(t\right)\right)\right]}
\end{equation}

is a weight function that depends on the normalization of the response.

\begin{figure*}[htbp]
    \centering
    \includegraphics[width=\textwidth]
    {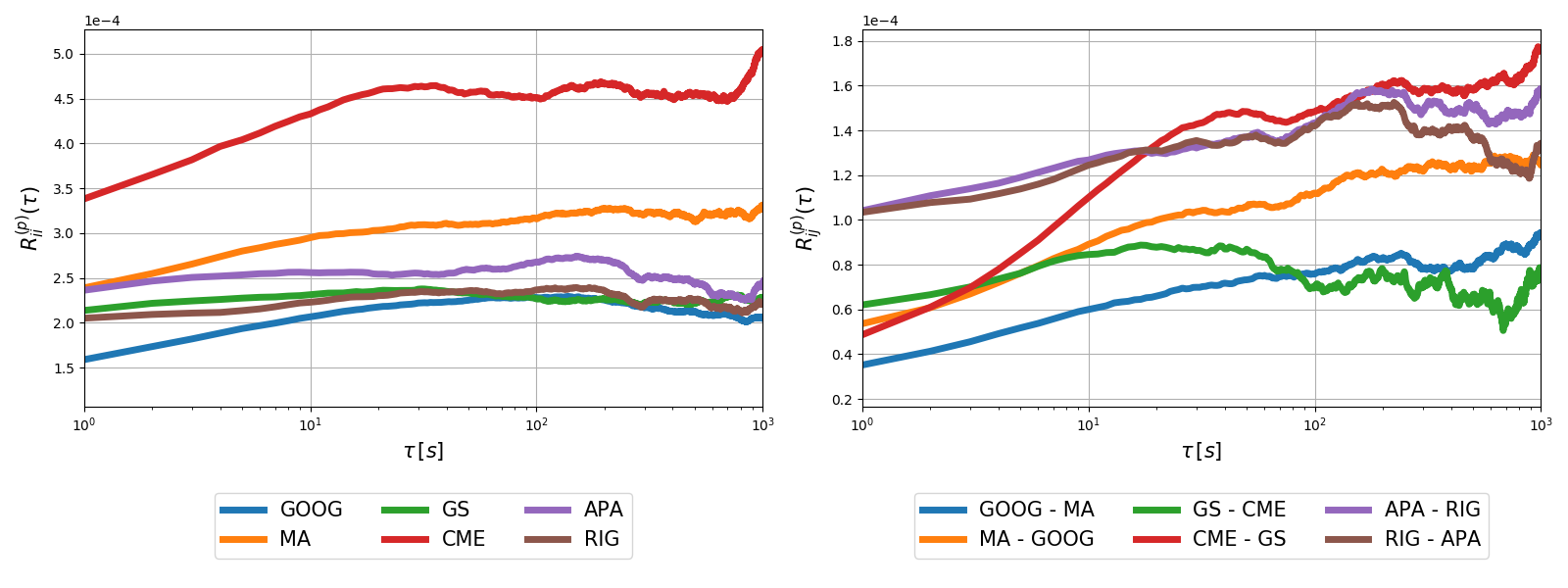}
    \caption{Self- and cross-response functions
             $R^{\left(p\right)}_{ij}\left(\tau\right)$ excluding
             $\varepsilon^{\left(p\right)}_{j}\left(t\right) = 0$ in 2008
             versus time lag $\tau$ on a logarithmic scale in physical time
             scale. Self-response functions (left) of individual stocks and
             cross-response functions (right) of stock pairs from the same
             economic sector.}
    \label{fig:market_response_time_scale}
\end{figure*}

The results showed in Fig. \ref{fig:market_response_time_scale} are the
self- and cross-response functions in physical time scale. For the
self-response functions we can say again that in almost all the cases, an
increase to a maximum is followed by a decrease. Thus, the trend in the self-
and cross-response is eventually reversed.
In the cross-response functions, we have a similar behavior with the previous
subsection, where the time lag in some pairs was not enough to see the decrease
of the response.

Compared with the response functions in trade time scale, the response functions
in physical time scale are stronger.

\subsection{Activity response functions on physical time scale}
\label{subsec:activity_response_function}

Finally, we define the activity self- and cross-response functions in physical
time scale, using the trade signs and the returns in physical time scale.
We add a factor $N_{j,d} \left(t \right)$ to check the influence of the
frequency of trades in a second in the response functions. The activity price
response function is defined as

\begin{equation}\label{eq:activity_response_functions_general}
    R^{\left(p, a\right)}_{ij}\left(\tau\right)=\left\langle r^{\left(p\right)}
    _{i}\left(t-1, \tau\right) \varepsilon_{j}^{\left(p\right)} \left(t\right)
    N \left(t \right) \right\rangle _{P}
\end{equation}

where the superscript $a$ refers to the activity response function. The
corresponding explicit expression reads

\begin{align}
    R_{ij}^{\left(p, a\right)}\left(\tau\right)&=\frac{1}{\sum_{d=d_{0}}
    ^{d_{f}} \sum_{t=t_{0}}^{t_{f}}N_{j,d} \left(t\right)} \nonumber \\
    &\sum_{d=d_{0}}^{d_{f}}\sum_{t=t_{0}}^{t_{f}}r^{\left(p\right)}_{i,d}
    \left(t-1,\tau\right) \varepsilon_{j,d}^{\left(p\right)}\left(t\right)
    N_{j,d} \left(t\right)\\
    &=\sum_{d=d_{0}}^{d_{f}} \sum_{t=t_{0}}^{t_{f}}r^{\left(p\right)}_{i,d}
    \left(t-1,\tau\right) \frac{\varepsilon_{j,d}^{\left(p\right)}\left(t \right)
    N_{j,d}\left(t\right)} {\sum_{d=d_{0}}^{d_{f}}\sum_{t=t_{0}}^{t_{f}}
    N_{j,d}\left(t \right)} \nonumber \\
    &=\sum_{d=d_{0}}^{d_{f}} \sum_{t=t_{0}}^{t_{f}}r^{\left(p\right)}_{i,d}
    \left(t-1,\tau\right) \text{sgn}\left(E_{j,d}\left(t\right)\right)
    w_{j,d}^{\left(a\right)}\left(t\right)
\end{align}
where
\begin{equation}
    w_{j,d}^{\left(a\right)}\left(t\right) = \frac{N_{j,d}\left(t \right)}
    {\sum_{d=d_{0}}^{d_{f}}\sum_{t=t_{0}}^{t_{f}}N_{j,d}\left(t\right)}
\end{equation}
is a weight function that depends on the normalization of the response.

As $E_{j,d}\left(t\right)$ is the sum of $+1$ and $-1$ in one second and
$N_{j,d}\left(t\right)$ is the number of trades in a second,
$N_{j,d}\left(t\right) \ge E_{j,d}\left(t\right)$.
$N_{j,d}\left(t\right) = E_{j,d}\left(t\right)$ only when all the trades in a
second are buys $(+1)$.

The trade weight $w_{j,d}^{\left(t\right)}\left(t\right)$ reduces noises, The
physical weight $w_{j,d}^{\left(p\right)}\left(t\right)$ gives every step the
same weight, and the activity weight $w_{j,d}^{\left(a\right)}\left(t\right)$
emphasizes seconds with large activity.

\begin{figure*}[htbp]
    \centering
    \includegraphics[width=\textwidth]
    {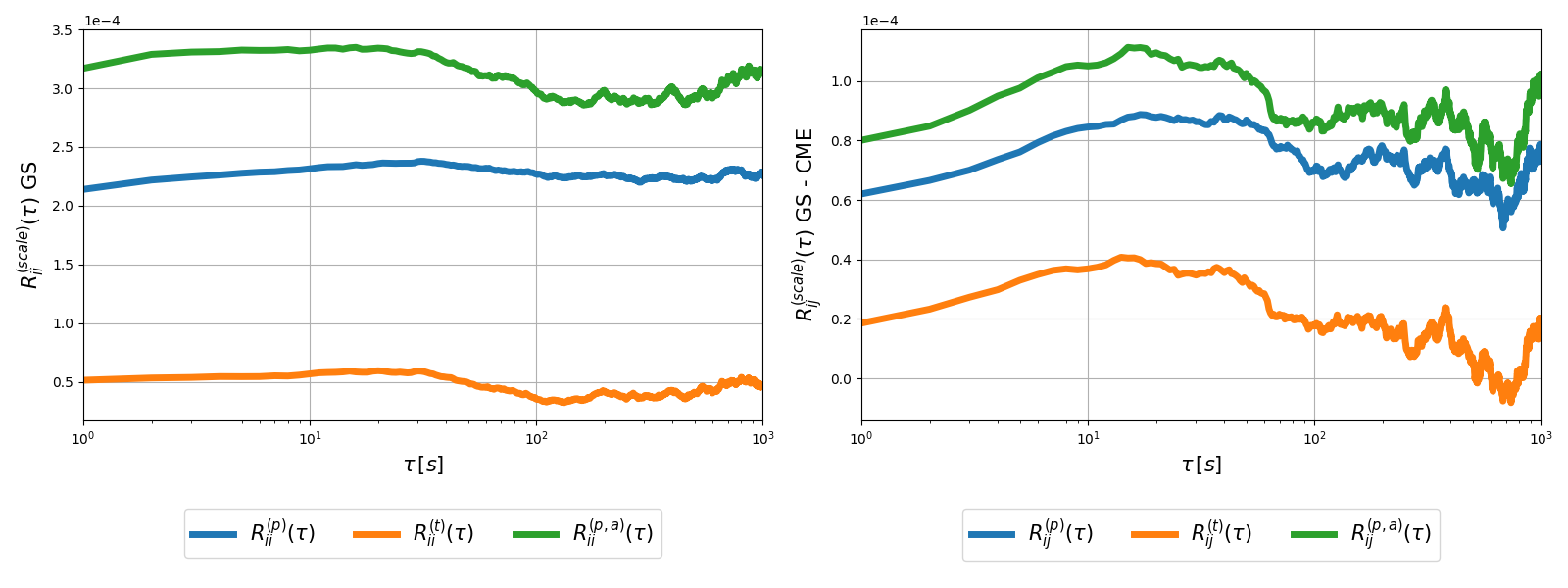}
    \caption{Self- and cross-response functions
             $R^{\left(scale\right)}_{ij}\left(\tau\right)$ excluding
             $\varepsilon^{\left(p\right)}_{j}\left(t\right) = 0$ in 2008
             versus time lag $\tau$ on a logarithmic scale. Self-response
             functions (left) of Goldman Sachs Group Inc. stock and
             cross-response functions (right) of Goldman Sachs Group Inc.-CME
             Group Inc. stocks.}
    \label{fig:relation_responses}
\end{figure*}

In Fig. \ref{fig:relation_responses}, we can see how the three
responses have approximately the same shape, but the strength of the signal
varies depending on the definition. The frequency of trades have a large
influence in the responses.

As predicted by the weights, the event response is weaker than the physical
response, and the activity response is the strongest response.

We propose a methodology to directly compare price response functions in trade
time scale and physical time scale. Additionally, we suggest a new definition
to measure the impact of the number of trades in physical time scale. In the
three curves in the figure can be seen the increase-decrease behavior of the
response functions.

Our results are consistent with the current literature, where the results
differ about a factor of two depending on the time scale. We note that the
activity response function implementation is only a test and was never defined
in previous works. That is why the difference of a factor of two can not be
seen in Fig. \ref{fig:relation_responses} for
$R_{ij}^{\left(p,a\right)} \left( \tau \right)$.

%% file: sections/07_time_shift.tex
\section{Time shift response functions}\label{sec:time_shift}

The relative position between returns and trade signs directly impact the
result of the response functions. Shifting the values to the right or to the
left either in trade time scale or physical time scale have approximately the
same effect.

To test this claim, we used the definition of the response function from Ref.
\cite{Wang_2016_cross} and add a parameter $t_{s}$ that shifts the position
between returns and trade signs. To see the impact of the time shift we
analyzed the stocks showed in Table \ref{tab:companies} in the year 2008. We
used different time shifts in the response function
\begin{equation}\label{eq:time_shift_general}
    R_{ij}^{\left(scale, s\right)}\left(\tau\right)=\left\langle
    r^{\left(scale\right)}_{i} \left(t-t_{s},\tau\right)
    \varepsilon^{\left(scale\right)}_{j} \left(t\right)\right\rangle_{average}
\end{equation}
where the index $i$ and $j$ correspond to stocks in the market,
$r^{\left(scale\right)}_{i}$ is the return of the stock $i$ in a time lag
$\tau$ with a time shift $t_{s}$ in the corresponding scale and
$\varepsilon^{\left(scale\right)}_{j}$ is the trade sign of the stock $j$ in
the corresponding scale.  The superscript $scale$ refers to the time scale
used, whether physical time scale ($scale = p$) or trade time scale
($scale = t$). $R_{ij}^{\left(scale, s\right)}$ is the time shift price
response function, where the superscript $s$ refers to the time shift. Finally,
The subscript $average$ refers to the way to average the price response,
whether relative to the physical time scale ($average = P$) or relative to the
trade time scale ($average = T$).

We compute the response functions according to two cases. In one case we set
$\tau$ to a constant value and vary $t_{s}$, and in the other case we set
$t_{s}$ to a constant value and vary $\tau$.

In Sect. \ref{subsec:time_shift_trade} we analyze the influence of the time
shift between the trade signs and returns in trade time scale and in Sect.
\ref{subsec:time_shift_physical} we analyze the influence of the time shift
between the trade signs and returns in physical time scale.

\subsection{Trade time scale shift response functions}
\label{subsec:time_shift_trade}

On the trade time scale we compute the response function

\begin{equation}\label{eq:time_shift_trade}
    R_{ij}^{\left(t, s\right)}\left(\tau\right)=\left\langle r^{\left(t\right)}
    _{i} \left(t-t_{s},\tau\right) \varepsilon^{\left(t\right)}_{j}
    \left(t\right)\right\rangle _{t}
\end{equation}

In this case for $r^{\left(t\right)}_{i}$, we associate all the trade signs to
a return value and create pseudo midpoint price values in trade time scale.
Then, we shift the trade signs and the returns by trades. Hence, the time lag
and time shift are in trade time scale.

\begin{figure*}[htbp]
    \centering
    \includegraphics[width=\textwidth]{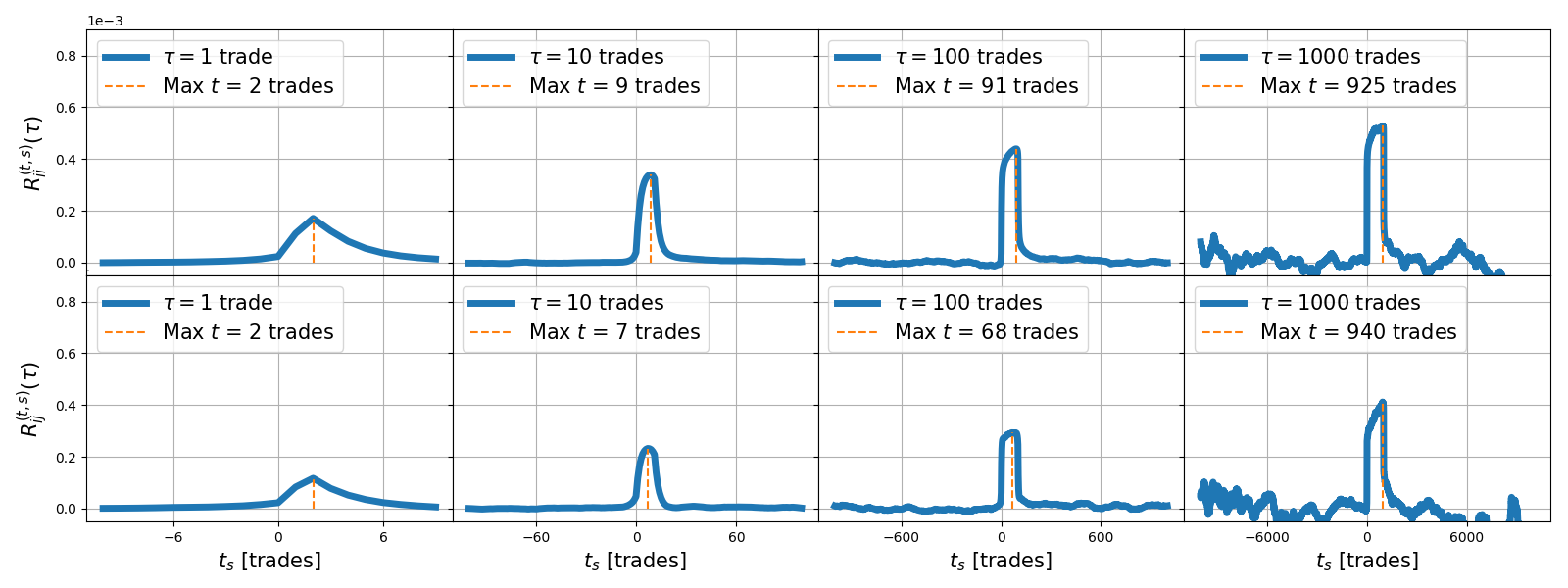}
    \caption{Self-response functions
             $R_{ii}^{\left(t, s\right)}\left(\tau\right)$ in 2008 versus shift
             for the Transocean Ltd. stock (top) and cross-response functions
             $R_{ij}^{\left(t, s\right)}\left(\tau\right)$ in 2008 versus shift
             for the Transocean Ltd.-Apache Corp. stocks (bottom) in trade time
             scale.}
    \label{fig:shift_trade_scale}
\end{figure*}

\begin{figure*}[htbp]
    \centering
    \includegraphics[width=\textwidth]{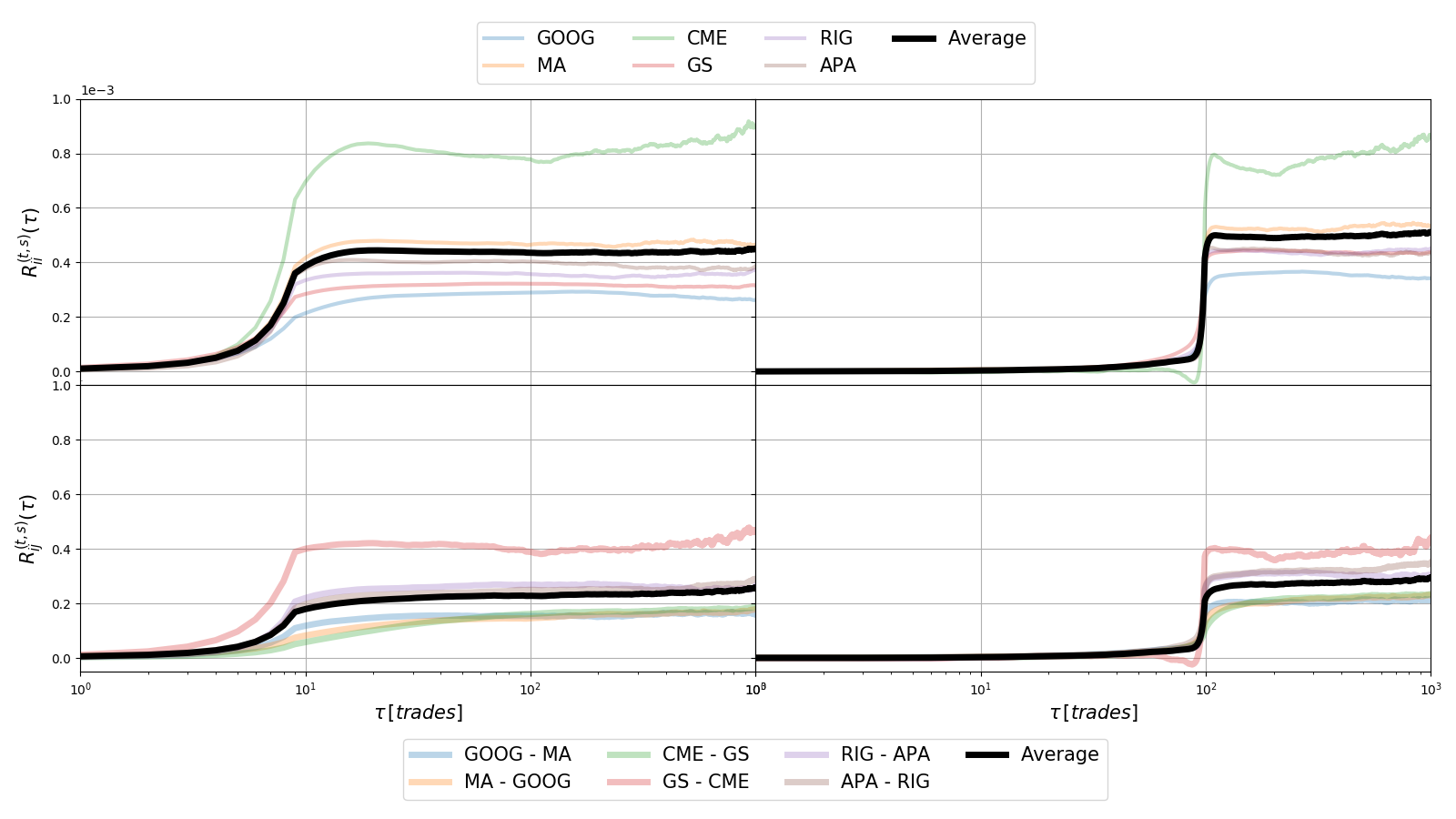}
    \caption{Self- and cross-response functions
             $R^{\left(t, s\right)}_{ij}\left(\tau\right)$ in 2008 versus time
             lag $\tau$ on a logarithmic scale for different shifts in trade
             time scale. Self-response functions (top) of individual stocks and
             cross-response functions (bottom) of stocks pairs from the same
             economic sector. We use time lag values $\tau=10$ trades (left)
             and $\tau=100$ (right).}
    \label{fig:shift_responses_trade_scale}
\end{figure*}

In Fig. \ref{fig:shift_trade_scale}, we show the response functions results for
fixed $\tau$ values while $t_{s}$ is variable. In the different $\tau$ values
figures, the results are almost the same. The response functions are zero
either if the time shift is larger than $\tau$, or if the time shift is smaller
than zero. However, related to the time lag, there is a zone where the signal
is different from zero. For values between zero and $\tau$ there is a peak in a
position related to $\tau$. The response function grows and decreases
relatively fast.

We tested the response function for fixed time shift values while $\tau$ is
variable. In Fig. \ref{fig:shift_responses_trade_scale} we use a time shift of
$10$ trades (left) and $100$ trades (right). In both, self- and cross-response
results are qualitatively the same. It can be seen that the response functions
have a zero signal before the time shift. After the returns and trade signs
find their corresponding order the signals grow. In comparison with the values
obtained in Fig. \ref{fig:response_function_trade_scale}, it looks like the
response function values with large time shift are stronger. However, this is
an effect of the averaging of the functions. As the returns and trade signs are
shifted, there are less values to average, and then the signals are stronger.
Anyway, the figure shows the importance of the position order between the trade
signs and returns to compute the response function.

\subsection{Physical time scale shift response functions}
\label{subsec:time_shift_physical}

In the physical time scale we compute the response function

\begin{equation}\label{eq:time_shift_physical}
    R_{ij}^{\left(p, s\right)}\left(\tau\right)=\left\langle r^{\left(p\right)}
    _{i} \left(t-t_{s},\tau\right) \varepsilon^{\left(p\right)}_{j}
    \left(t\right)\right\rangle _{t}
\end{equation}

\begin{figure*}[htbp]
    \centering
    \includegraphics[width=\textwidth]{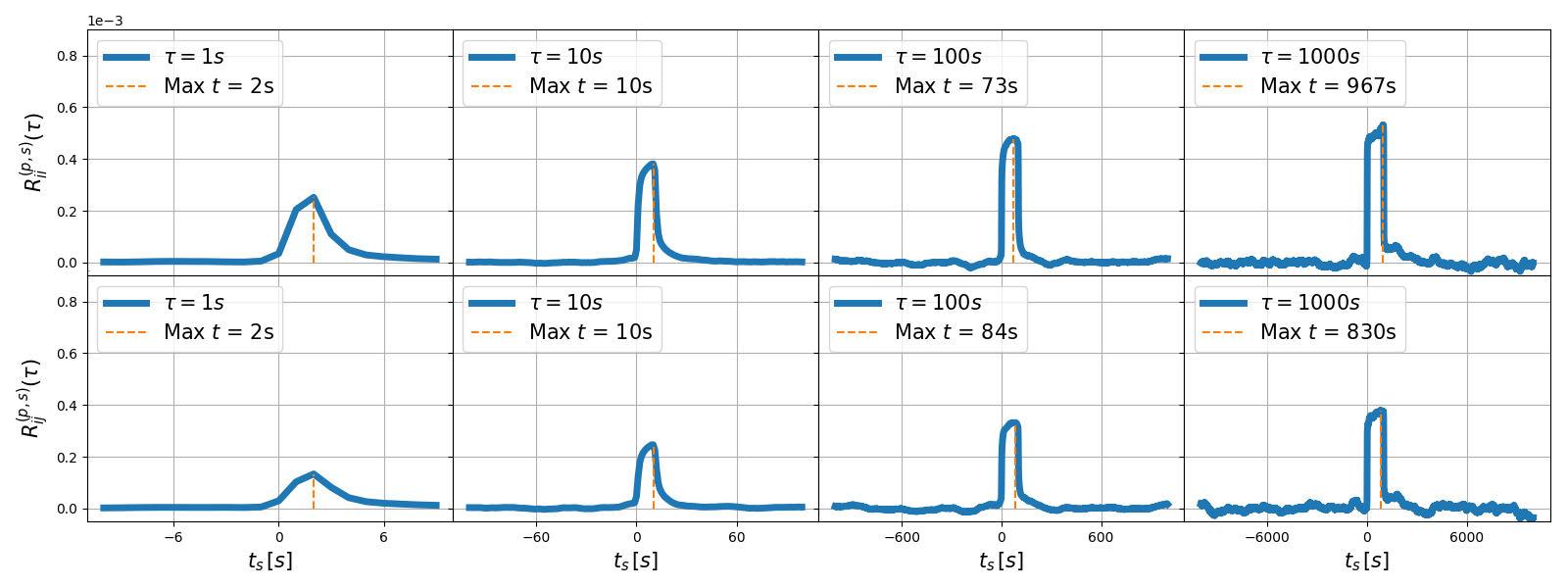}
    \caption{Self-response functions
             $R_{ii}^{\left(p, s\right)}\left(\tau\right)$ excluding
             $\varepsilon^{\left(p\right)}_{i}\left(t\right) = 0$ in 2008
             versus shift for the Transocean Ltd. stock (top) and
             cross-response functions
             $R_{ij}^{\left(p, s\right)}\left(\tau\right)$ excluding
             $\varepsilon^{\left(p\right)}_{j}\left(t\right) = 0$ in 2008
             versus shift for the Transocean Ltd.-Apache Corp. stocks (bottom)
             in physical time scale.}
    \label{fig:shift_physical_scale}
\end{figure*}

\begin{figure*}[htbp]
    \centering
    \includegraphics[width=\textwidth]{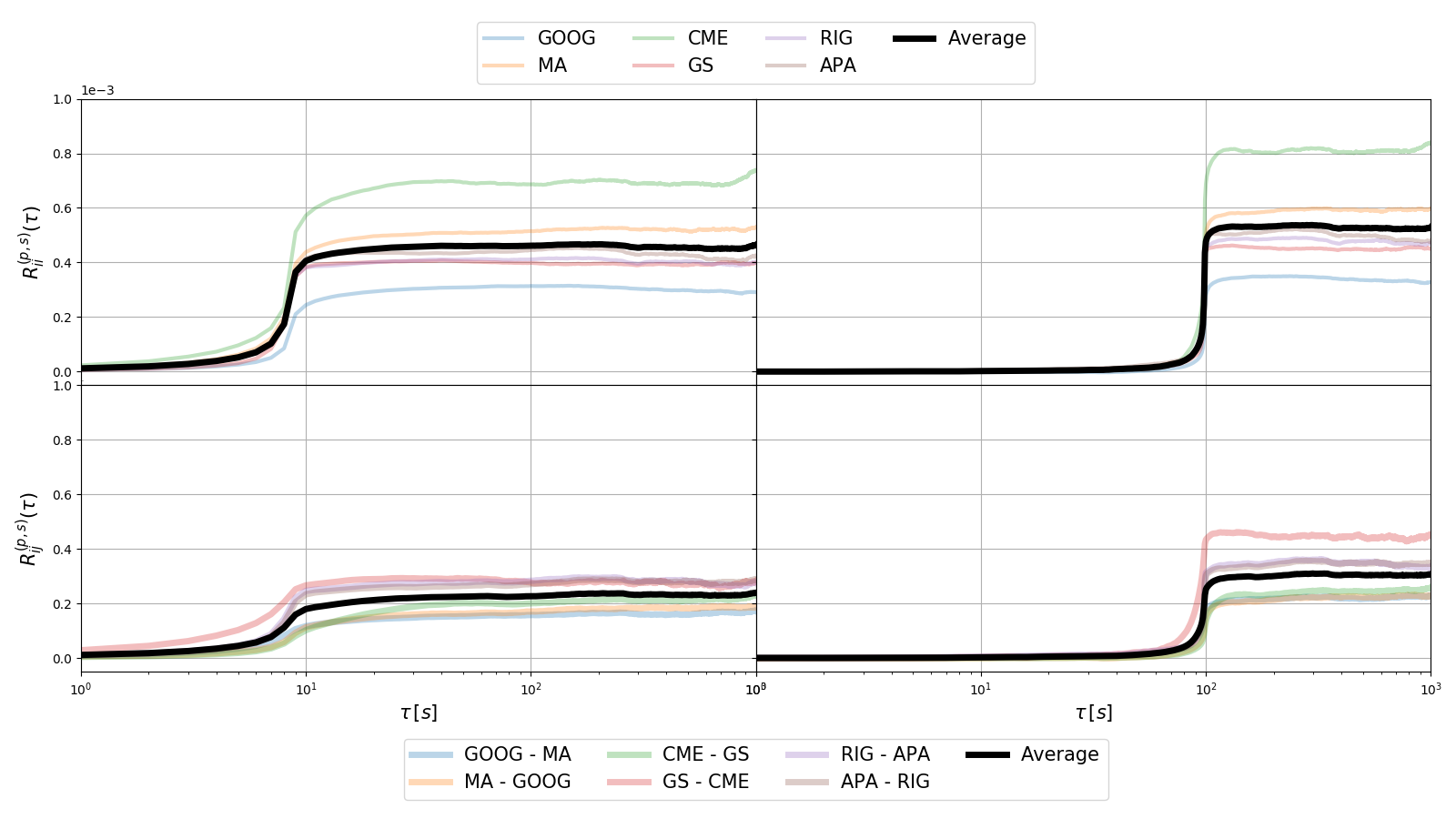}
    \caption{Self- and cross-response functions
             $R^{\left(p, s\right)}_{ij}\left(\tau\right)$ excluding
             $\varepsilon^{\left(p\right)}_{j}\left(t\right) = 0$ in 2008
             versus time lag $\tau$ on a logarithmic scale for different shifts
             in physical time scale. Self-responses functions (top) of
             individual stocks and cross-response functions (bottom) of stocks
             pairs from the same economic sector. We use time lag values
             $\tau=10$ trades (left) and $\tau=100$ (right).}
    \label{fig:shift_responses_physical_scale}
\end{figure*}

Similar to the results in Subsect. \ref{subsec:time_shift_trade}, Fig.
\ref{fig:shift_physical_scale} shows the responses functions for fixed $\tau$
values while $t_{s}$ is variable. Again, the response functions are zero if the
time shift is larger than the time lag, or if the time shift is smaller than
zero. For every $\tau$ value, there is a peak. The peak grows and decay
relatively fast. The response signal usually starts to grow in zero or a little
bit earlier and grows to a value around to $\tau$. In this zone the response
functions are different to zero.

The results for fixed time shift values and variable time lag ($t_{s}=10s$ and
$t_{s}=100s$) are shown in Fig. \ref{fig:shift_responses_physical_scale}. The
self- and cross-response results are qualitatively the same compared with the
previous subsection. The response functions are zero before the time shift
value. After the returns and the trade signs reach their order, the signals
grow. The same effect of the apparent stronger signal can be seen here, and
again, it is due to the averaging values.

The results in trade time scale and physical time scale can be explained
understanding the dynamics of the market. A trade can or can not change the
price of a ticker. Therefore, when a change in price happens, a change in
midpoint price, and consequently in returns happens. Thus, it is extremely
important to keep the order of the events and the relation between them. When
we shift the trade signs and returns, this order is temporarily lost and as
outcome the signal does not have any meaningful information. When the order is
recovered during the shift, the signal grows again, showing response function
values different to zero. In this section we were interested only in the order
(shift) and not in the responses values, which were analyzed in Sect.
\ref{sec:response_functions_imp}. A time shift smaller than zero does not have
any useful information about the response. If the time shift is equal to zero,
the signal is weak, due to the time needed by the market to react to the new
information. On the other hand, a time shift larger than two steps shows the
information is lost and the signal only grows when the original order is
resumed.

Then the question is what is the ideal time shift to compute the response
functions. Our approach in Sect. \ref{sec:response_functions_imp} takes in
account that the changes in the quotes are the ones that attract the agents to
buy or sell their shares. Hence, they directly impact the trade signs.
According to the results, the response can take up to two time steps in the
corresponding scale to react to the change in quotes. Thus, a time shift
larger than two time steps makes no sense.
On the other hand, in the case of the physical time scale, where a sampling is
used, to assure the selection of a midpoint price at the beginning of a second,
it is a good strategy to use the last midpoint of the previous second as the
first midpoint price of the current second. In this case an apparent one second
shift is used between returns and trade signs.

%% file: sections/08_short_long.tex
\section{Time lag analysis}\label{sec:short_long}

\begin{figure}[htbp]
    \centering
    \includegraphics[width=\columnwidth]{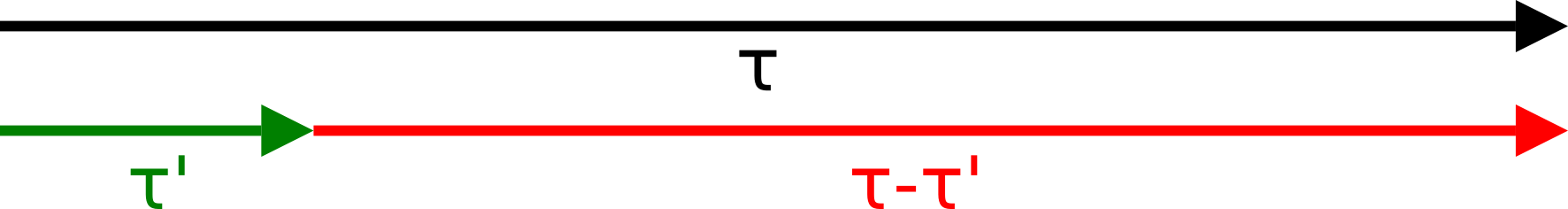}
    \caption{$\tau$ value divided in short and long time lag.}
    \label{fig:tau_short_long}
\end{figure}

Regarding Equation (\ref{eq:return_general}), we use a time lag $\tau$ in the
returns to see the gains or loses in a future time. However, the strength of
the return in the time lag should not be equal along its length. Then, we
divide the full range time lag $\tau$ in an immediate time lag and in a late
time lag as show in Fig. \ref{fig:tau_short_long}, where

\begin{equation}\label{eq:tau_short_long}
    \tau = \tau' + \left( \tau - \tau' \right)
\end{equation}
for $\tau' < \tau$. This distinguish the returns depending on the time lag as
the short (immediate) return $\tau'$ with the long return $\tau - \tau'$.
This approach is similar to the concept used in Ref. \cite{Wang_2018_b}, where
the price impact for a single trade is estimated by the immediate response on
an event time scale. In our case, we check all over the range of the time lag
in the price response function on a physical time scale.

To use the short and long time lag, we rewrite the returns in physical time
scale as

\begin{align}\label{eq:short_long_return}
    r^{\left(p, sl\right)}_{i}\left(t,\tau\right)&=\ln\left(\frac{m_{i}\left(t+\tau\right)}
    {m_{i} \left(t\right)}\right) \nonumber \\
    &=\ln\left[\left(\frac{m_{i}\left(t+\tau\right)}{m_{i}\left(t+\tau'\right)}
    \right)
    \left(\frac{m_{i} \left(t+\tau'\right)}{m_{i}\left(t\right)}\right)\right]
    \nonumber \\
    &=\ln\left(\frac{m_{i}\left(t+\tau\right)}{m_{i}\left(t+\tau'\right)}
    \right)+ \ln\left(\frac{m_{i}\left(t+\tau'\right)}{m_{i}\left(t\right)}
    \right)\nonumber \\
    &\approx\frac{m_{i}\left(t+\tau\right)-m_{i}\left(t+\tau'\right)}
    {m_{i}\left(t+\tau'\right)} +\frac{m_{i}\left(t+\tau'\right)-m_{i}
    \left(t\right)}{m_{i}\left(t\right)}
\end{align}

where the superscript $sl$ refers to short-long and the second term of the
right part is constant with respect to $\tau$. Replacing Eq.
(\ref{eq:short_long_return}) in the price response function in physical time
scale (Eq. (\ref{eq:response_functions_time_scale_general})) we have

\begin{align}\label{eq:short_long_response}
    R^{\left(p, sl\right)}_{ij}\left(\tau\right)&=\left\langle
    r^{\left(p, sl\right)}_{i}\left(t - 1, \tau\right)
    \varepsilon^{\left(p\right)}_{j}\left(t\right)\right\rangle _{P}\nonumber\\
    &\approx\left\langle \frac{m_{i}\left(t - 1 +\tau\right)-m_{i}
    \left(t - 1 +\tau'\right)} {m_{i}\left(t - 1 +\tau'\right)}
    \varepsilon^{\left(p\right)}_{j} \left(t\right)\right\rangle_{P}\nonumber\\
    & +\left\langle \frac{m_{i} \left(t - 1 +\tau'\right)-m_{i}
    \left(t - 1\right)}{m_{i}\left(t - 1\right)}
    \varepsilon^{\left(p\right)}_{j}\left(t\right)\right\rangle _{P}
\end{align}

Where the first term in the right side of Eq. (\ref{eq:short_long_response})
is the long response and the right term is the short response. Again, the right
term of Eq. (\ref{eq:short_long_response}) is independent of $\tau$.

\begin{figure*}[htbp]
    \centering
    \includegraphics[width=\textwidth]
    {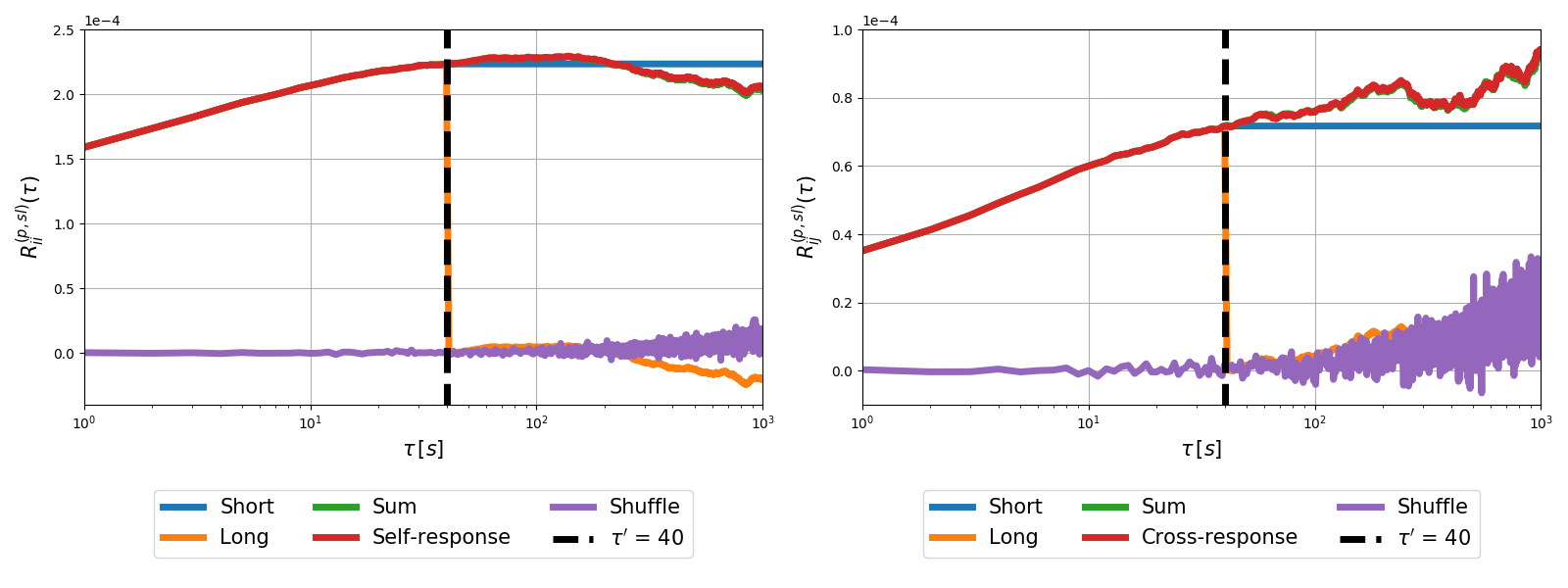}
    \caption{Self- and cross-response functions
             $R^{\left(p, sl\right)}_{ij}\left(\tau\right)$ excluding
             $\varepsilon^{\left(p\right)}_{j}\left(t\right) = 0$ in 2008
             versus time lag $\tau$ on a logarithmic scale using a $\tau'=40$
             in physical time scale. Self-response functions (left) of Alphabet
             Inc. stock and cross-response functions (right) of Alphabet
             Inc.-Mastercard Inc. stocks.}
    \label{fig:short_long_responses}
\end{figure*}

The results in Fig. \ref{fig:short_long_responses} show the short response, the
long response, the addition of the short response and long response (Sum), the
original response, a random response and the value of $\tau'$.

The main signal of the response function come from the short response.
Depending on the stock and the value of $\tau'$ the long response can increase
or decrease the short response signal, but in general the long response does
not give a significant contribution to the complete response.

Before $\tau'$, the short response and long response are the same, as the self
and cross-response definition do not define values smaller than $\tau '$, so it
is computed as the original response. In the figure, the curves of the short
and long response are under the curve of the original response. After $\tau'$,
the short response is a strong constant signal. On the other hand, the long
response immediately fades, showing the small contribution to the final
response. To compare the significance of the long response, We added a random
response made with the trade signs used to compute the response but with a
shuffle order. The long response and the random response are comparable, and
show how the long response is not that representative in the final response.
If we add the short and long response, we obtain the original response. In Fig.
\ref{fig:short_long_responses}, the original response (red line) has the same
shape to the addition of the short and long response (green line).

For the response functions that show the increase-decrease behavior in between
the time lag $\tau = 10^{3}$, the peak is usually between $\tau = 10^{1}$ and
$\tau = 10^{2}$. In these cases the long response are always negative after the
$\tau'$ value and is comparable in magnitude with the random signal.
On the other hand, the response functions that requires a bigger time lag to
show the increase-decrease behavior, have non negative long responses, but
still they are comparable in magnitude with the random signal.

According to our results, price response functions show a large impact on the
first instants of the time lag. We proposed a new methodology to measure this
effect and evaluate their consequences. The short response dominates the
signal, and the long response vanishes.

%% file: sections/09_spread_impact.tex
\section{Spread impact in price response functions}\label{sec:spread_impact}

When we calculate the price response functions, the signal of the response
depends directly on the analyzed stock. Thus, even if the responses functions
are in the same scale, their values differ from one to another. We choose the
spread \cite{reg_and_irreg} to group 524 stocks in the NASDAQ stock market for
the year 2008 in physical time scale, and check how the average strength of the
price self-response functions in physical time scale behaved for this groups.
For each stock we compute the spread in every second along the market time.
Then we average the spread during the 253 business days in 2008. With this
value we group the stocks.

We used three intervals to select the stocks groups ($s<0.05\$$,
$0.05\$ \le s <0.10\$$ and $0.10\$ \le s <0.40\$$). The detailed information of
stocks, the spread and the groups can be seen in Appendix
\ref{app:spread_impact}. With the groups of the stocks defined, we averaged the
price response functions of each group.

\begin{figure}[htbp]
    \centering
    \includegraphics[width=\columnwidth]{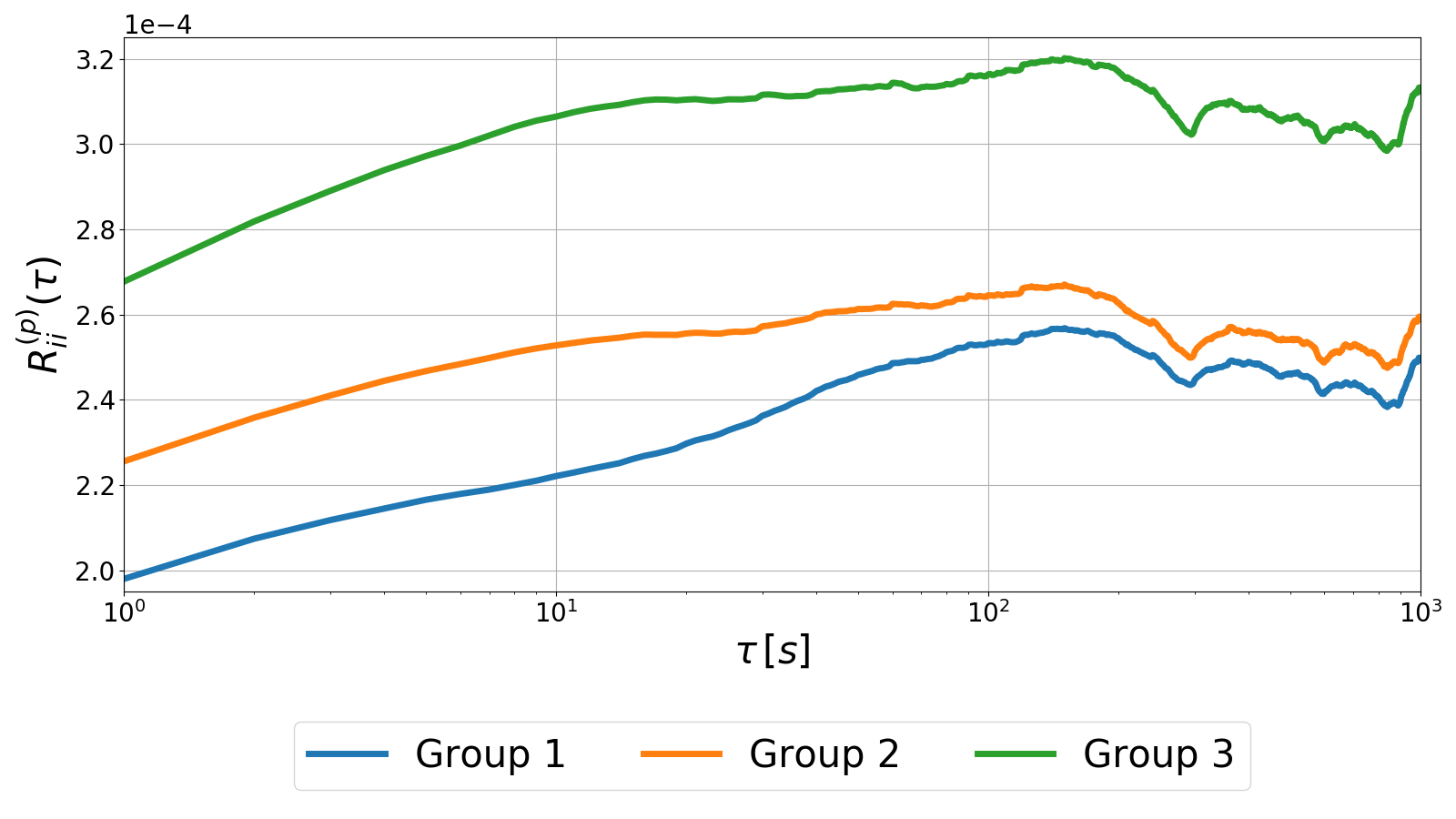}
    \caption{Average price self-response functions
             $R^{\left(p\right)}_{ii}\left(\tau\right)$ excluding
             $\varepsilon^{\left(p\right)}_{i}\left(t\right) = 0$ in 2008
             versus time lag $\tau$ on a logarithmic scale in physical time
             scale for 524 stocks divided in three representative groups.}
    \label{fig:spread_impact}
\end{figure}

In Fig. \ref{fig:spread_impact} we show the average response functions for
the three groups. The average price response function for the stocks with
smaller spreads (more liquid) have in average the weakest signal in the figure.
On the other hand, the average price response function for the stocks with
larger spreads (less liquid) have in average the strongest signal. According to
the results described in Sect. \ref{sec:response_functions_def} and
\ref{sec:response_functions_imp}, the average price response functions for all
the groups follow the increment to a maximum followed by a decrease in the
signal intensity.

The strength of the price self-response function signals grouped by the spread
can be explained knowing that the response functions directly depend on the
trade signs. As long as the stock is liquid, the number of trade signs grow.
Thus, at the moment of the averaging, the large amount of trades, reduces the
response function signal. Therefore, the response function decrease as long as
the liquidity grows, and as stated in the introduction the spread is negatively
related to trading volume, hence, firms with more liquidity tend to have lower
spreads.

Finally, an interesting behavior can be seen in Fig. \ref{fig:spread_impact}.
Despite each stock has a particular price response according to the returns and
trade signs, the average price response function for the different groups
seems to be quite similar. As all the analyzed stocks come from the same market
we can infer that the general behavior of the market affect all the stocks,
influencing in average a group response.

%% file: sections/10_conclusion.tex
\section{Conclusion}\label{sec:conclusion}

We went into detail about the response functions in correlated financial
markets. We define the trade time scale and physical time scale to compute the
self- and cross-response functions for six companies with the largest average
market capitalization for three different economic sectors of the S\&P index
in 2008.
Due to the characteristics of the data used, we had to classify and sampling
values to obtain the corresponding quantities in different time scales.
The classification and sampling of the data had impact on the results, making
them smoother or stronger, but always keeping their shape and behavior.

The response functions were analyzed according to the time scales. We proposed
a new approach to compare price response functions from different scales. We
used the same midpoint prices in physical time scale with the corresponding
trade signs in trade time scale or physical time scale. This assumption allowed
us to compare both price response functions and get an idea of how
representative was the behavior obtained in both cases. For trade time scale,
the signal is weaker due to the large averaging values from all the trades in a
year. In the physical time scale, the response functions had less noise and
their signal were stronger. We proposed an activity response to measure how the
number of trades in every second highly impact the responses. As the response
functions can not grow indefinitely with the time lag, they increase to a peak,
to then decrease. It can be seen that the market needs time to react and revert
the growing. In both time scale cases depending on the stocks, two
characteristics behavior were shown. In one, the time lag was large enough to
show the complete increase-decrease behavior. In the other case, the time lag
was not enough, so some stocks only showed the growing behavior.

We modify the response function to add a time shift parameter. With this
parameter we wanted to analyze the importance in the order of the relation
between returns and trade signs. In trade time scale and physical time scale we
found similar results. When we shift the order between returns and trade signs,
the information from the relation between them is temporarily lost and as
outcome the signal does not have any meaningful information. When the order is
recovered, the response function grows again, showing the expected shape.
We showed that this is not an isolated conduct, and that all the shares used in
our analysis exhibit the same behavior. Thus, even if they are values of time
shift that can give a response function signal, empirically we propose this
time shift should be a value between $t_{s} = \left(0,2\right]$ time steps.

We analyzed the impact of the time lag in the response functions. We divided
the time lag in a short and long time lag. With this division we adapted the
price response function in physical time scale. The response function that
depended on the short time lag, showed a stronger response. The long response
function vanish, and depending on the stock could take negative and
non-negative values comparable to a random signal.

Finally, we checked the spread impact in price self-response functions. We
divided 524 stocks from the NASDAQ stock market in three groups depending on
the year average spread of every stock. The response functions signal were
stronger for the group of stocks with the larger spreads and weaker for the
group of stocks with the smaller spreads. A general average price response
behavior was spotted for the three groups, suggesting a market effect on the
stocks.

%% file: sections/11_paper_contributions.tex
\section{Author contribution statement}

TG proposed the research. SMK and JCHL developed the method of analysis.
The idea to look the time shift and to analyze the spread impact was due to
JCHL, and the idea of the time lag analysis was due to SMK. JCHL carried out
the analysis. All the authors contributed equally to analyze the results and
write the paper.

\begin{acknowledgement}
    We thank S. Wang for fruitful discussions.
    One of us (JCHL) acknowledges financial support from the German
    Academic Exchange Service (DAAD) with
    the program ``Research Grants - Doctoral Programmes in Germany''
    (Funding programme 57381412)
\end{acknowledgement}

%% file: sections/12_appendix_A.tex
\section{NASDAQ stocks used to analyze the spread impact}
\label{app:spread_impact}

We analyzed the spread impact in the response functions for 524 stocks from the
NASDAQ stock market for the year 2008. In Tables \ref{tab:spread_comp_1},
\ref{tab:spread_comp_2}, \ref{tab:spread_comp_3} and \ref{tab:spread_comp_4},
we listed the stocks in their corresponding spread groups.

\afterpage{
    \clearpage
    \begin{landscape}
    \begin{table}
    \begin{threeparttable}
    \caption{Information of the stocks in Group 1.}

    \begin{tabular}{lllllllllll}
    \multicolumn{3}{c}{Group 1} &  & Symbol & Company & Spread\tnote{1} &  & Symbol & Company & Spread\tnote{1}\tabularnewline
    \cline{1-3} \cline{5-7} \cline{9-11}
    Symbol & Company & Spread\tnote{1} &  & LUV & Southwest Airlines Company & $0.01\$$ &  & TSN & Tyson Foods Inc. & $0.02\$$\tabularnewline
    \cline{1-3}
    F & Ford Motor Company & $0.01\$$ &  & BRCM & Broadcom Inc. & $0.01\$$ &  & CA & CA Inc. & $0.02\$$\tabularnewline
    Q & Qwest Communications Int. & $0.01\$$ &  & TXN & Texas Instruments Inc. & $0.01\$$ &  & XEL & Xcel Energy Inc. & $0.02\$$\tabularnewline
    ETFC & E-Trade Financial Corp. & $0.01\$$ &  & TER & Teradyne Inc. & $0.01\$$ &  & AA & Alcoa Corp. & $0.02\$$\tabularnewline
    PFE & Pfizer Inc. & $0.01\$$ &  & MYL & Mylan N.V. & $0.01\$$ &  & KR & Kroger Company & $0.02\$$\tabularnewline
    MOT & Motus GI Holdings Inc. & $0.01\$$ &  & HCBK & Hudson City Bancorp & $0.01\$$ &  & MRK & Merck \& Company Inc. & $0.02\$$\tabularnewline
    AMD & Advanced Micro Devices & $0.01\$$ &  & SPLS & Staples Inc. & $0.01\$$ &  & NSM & Nationstar Mortgage Holdings & $0.02\$$\tabularnewline
    TLAB & Tellabs Inc. & $0.01\$$ &  & SGP & Siamgas and Petrochemicals & $0.01\$$ &  & WMT & Walmart Inc. & $0.02\$$\tabularnewline
    INTC & Intel Corp. & $0.01\$$ &  & HST & Host Hotels \& Resorts Inc. & $0.01\$$ &  & FITB & Fifth Third Bancorp & $0.02\$$\tabularnewline
    TWX & Time Warner Inc. & $0.01\$$ &  & AES & Aes Corp. & $0.01\$$ &  & EK & Eastman Kodak Company & $0.02\$$\tabularnewline
    CSCO & Cisco Systems Inc. & $0.01\$$ &  & KFT & Kraft Foods Inc. & $0.01\$$ &  & PHM & PulteGroup Inc. & $0.02\$$\tabularnewline
    THC & Tenet Healthcare Corp. & $0.01\$$ &  & NTAP & NetApp Inc. & $0.01\$$ &  & JPM & JP Morgan Chase \& Co. & $0.02\$$\tabularnewline
    LSI & Life Storage Inc. & $0.01\$$ &  & BAC & Bank of America Corp. & $0.01\$$ &  & WAG & Walgreen Co. & $0.02\$$\tabularnewline
    MU & Micron Technology Inc. & $0.01\$$ &  & HD & Home Depot Inc. & $0.01\$$ &  & SCHW & Charles Schwab Corp. & $0.02\$$\tabularnewline
    EMC & EMC Corp. & $0.01\$$ &  & SOV & Life Storage Inc. & $0.01\$$ &  & RF & Regions Financial Corp. & $0.02\$$\tabularnewline
    MSFT & Microsoft Corp. & $0.01\$$ &  & QLGC & QLogic Corp. & $0.01\$$ &  & ADBE & Adobe Inc. & $0.02\$$\tabularnewline
    NOVL & Novell Inc. & $0.01\$$ &  & T & AT\&T Inc. & $0.01\$$ &  & MAT & Mattel Inc. & $0.02\$$\tabularnewline
    JAVA & Sun Microsystems Inc. & $0.01\$$ &  & GPS & Gap Inc. & $0.01\$$ &  & PAYX & Paychex Inc. & $0.02\$$\tabularnewline
    ORCL & Oracle Corp. & $0.01\$$ &  & DIS & Walt Disney Company & $0.01\$$ &  & PGR & Progressive Corp. & $0.02\$$\tabularnewline
    S & Sprint Nextel Corp. & $0.01\$$ &  & GM & General Motors Company & $0.01\$$ &  & HPQ & HP Inc. & $0.02\$$\tabularnewline
    DELL & Dell Technologies Inc. & $0.01\$$ &  & JNPR & Juniper Networks Inc. & $0.01\$$ &  & DOW & Dow Inc. & $0.02\$$\tabularnewline
    AMAT & Applied Material Inc. & $0.01\$$ &  & LOW & Lowe's Companies Inc. & $0.01\$$ &  & TE & TECO Energy Inc. & $0.02\$$\tabularnewline
    SLE & Spark Energy Inc. & $0.01\$$ &  & CBS & CBS Corp. & $0.01\$$ &  & BBBY & Bed Bath \& Beyond Inc. & $0.02\$$\tabularnewline
    SBUX & Starbucks Corp. & $0.01\$$ &  & CAG & ConAgra Brands Inc. & $0.01\$$ &  & JNJ & Johnson \& Johnson & $0.02\$$\tabularnewline
    DYN & Dynergy Inc. & $0.01\$$ &  & LLTC & Linear Technology Corp. & $0.01\$$ &  & IP & International Paper Company & $0.02\$$\tabularnewline
    DUK & Duke Energy Corp. & $0.01\$$ &  & DTV & DirecTV Group  & $0.01\$$ &  & RSH & Respiri Ltd. & $0.02\$$\tabularnewline
    CMCSA & Comcast Corp. & $0.01\$$ &  & HBAN & Huntington Bancshares Inc. & $0.01\$$ &  & MER & Mears Group PLC & $0.02\$$\tabularnewline
    IPG & Interpublic Group of Co. & $0.01\$$ &  & KG & Kinross Gold Corp. & $0.01\$$ &  & HAL & Halliburton Company & $0.02\$$\tabularnewline
    BSX & Boston Scientific Corp. & $0.01\$$ &  & CMS & CMS Energy Corp. & $0.01\$$ &  & KO & Coca-Cola Company & $0.02\$$\tabularnewline
    GE & General Electric Company & $0.01\$$ &  & ODP & Office Depot Inc. & $0.01\$$ &  & PBCT & People's United Financial Inc. & $0.02\$$\tabularnewline
    SYMC & Symantec Corp. & $0.01\$$ &  & NVLS & Nivalis Therapeutics Inc. & $0.01\$$ &  & WU & Western Union Company & $0.02\$$\tabularnewline
    C & Citigroup Inc. & $0.01\$$ &  & WFC & Wells Fargo \& Company & $0.01\$$ &  & USB & U.S. Bancorp & $0.02\$$\tabularnewline
    CPWR & Ocean Thermal Energy & $0.01\$$ &  & MO & Altria Group Inc. & $0.01\$$ &  & MCHP & Microchip Technology Inc. & $0.02\$$\tabularnewline
    NVDA & Nvidia Corp. & $0.01\$$ &  & VZ & Verizon Communications & $0.01\$$ &  & SO & Southern Company & $0.02\$$\tabularnewline
    YHOO & Yahoo Inc. & $0.01\$$ &  & DHI & D. R. Horton Inc. & $0.01\$$ &  & BJS & BJ's Wholesale Club Holdings & $0.02\$$\tabularnewline
    BMY & Bristol-Myers Squibb Co. & $0.01\$$ &  & SNDK & Sandisk Corp. & $0.01\$$ &  & MAS & Masco Corp. & $0.02\$$\tabularnewline
    ALTR & Altair Engineering Inc. & $0.01\$$ &  & CCE & Coca-Cola Enterprises & $0.01\$$ &  & NWL & Newell Brands Inc. & $0.02\$$\tabularnewline
    GLW & Corning Inc. & $0.01\$$ &  & NI & NiSource Inc. & $0.01\$$ &  & M & Macy's Inc. & $0.02\$$\tabularnewline
    JDSU & JDS Uniphase Corp. & $0.01\$$ &  & EXPE & Expedia Group Inc. & $0.01\$$ &  & CVS & CVS Health Corp. & $0.02\$$\tabularnewline
    NCC & NCC Group & $0.01\$$ &  & AIG & American International & $0.01\$$ &  & CTSH & Cognizant Tech. Solutions & $0.02\$$\tabularnewline
    EBAY & eBay Inc. & $0.01\$$ &  & INTU & Intuit Inc. & $0.01\$$ &  & GNW & Genworth Financial Inc. & $0.02\$$\tabularnewline
    EP & El Paso Corp. & $0.01\$$ &  & LTD & Limited Brands Inc. & $0.01\$$ &  & DFS & Discover Financial Services & $0.02\$$\tabularnewline
    XRX & Xerox Holdings Corp. & $0.01\$$ &  & CNP & CenterPoint Energy Inc. & $0.02\$$ &  & KEY & KeyCorp & $0.02\$$\tabularnewline
    WIN & Windstream Holdings Inc. & $0.01\$$ &  & QCOM & Qualcomm Inc. & $0.02\$$ &  & ADI & Analog Devices Inc. & $0.02\$$\tabularnewline
    XLNX & Xilinx Inc. & $0.01\$$ &  & JBL & Jabil Inc. & $0.02\$$ &  & SYY & Sysco Corp. & $0.02\$$\tabularnewline
    \end{tabular}

    \label{tab:spread_comp_1}
    \begin{tablenotes}
    \item[1] Average spread from 9:40:00 to 15:50:00 New York time during 2008.
    \end{tablenotes}

    \end{threeparttable}
    \end{table}
    \end{landscape}
    \clearpage
}

\afterpage{
    \clearpage
    \begin{landscape}
    \begin{table}
    \begin{threeparttable}
    \caption{Information of the stocks in Group 1.}

    \begin{tabular}{lllllllllll}
    \multicolumn{3}{c}{Group 1} &  & Symbol & Company & Spread\tnote{1} &  & Symbol & Company & Spread\tnote{1}\tabularnewline
    \cline{1-3} \cline{5-7} \cline{9-11}
    Symbol & Company & Spread\tnote{1} &  & TGT & Target Corp. & $0.02\$$ &  & MI & Marshall and Lisley Corp. & $0.03\$$\tabularnewline
    \cline{1-3}
    WMB & Williams Companies Inc. & $0.02$\$ &  & VLO & Valero Energy Corp. & $0.02\$$ &  & AEP & American Electric Power Co. & $0.03\$$\tabularnewline
    SE & Sea Ltd American Dep. & $0.02\$$ &  & MMC & Marsh \& McLennan Co. & $0.02\$$ &  & NEM & Newmont Corp. & $0.03\$$\tabularnewline
    PG & Procter \& Gamble Co. & $0.02\$$ &  & CPB & Campbell Soup Company & $0.02\$$ &  & MRO & Marathon Oil Corp. & $0.03\$$\tabularnewline
    CIEN & Ciena Corp. & $0.02\$$ &  & TYC & Tyco International PLC & $0.02\$$ &  & ITW & Illionois Tool Works Inc. & $0.03\$$\tabularnewline
    LEN & Lennar Corp. & $0.02\$$ &  & MCD & McDonald's Corp. & $0.02\$$ &  & FFIV & F5 Networks Inc. & $0.03\$$\tabularnewline
    AKAM & Akamai Technologies & $0.02\$$ &  & HON & Honeywell International Inc. & $0.02\$$ &  & CVX & Chevron Corp. & $0.03\$$\tabularnewline
    UNH & UnitedHealth Group Inc. & $0.02\$$ &  & DF & Dean Foods Company & $0.02\$$ &  & PCAR & PACCAR Inc. & $0.03\$$\tabularnewline
    DD & DuPont de Nemours Inc. & $0.02\$$ &  & NWSA & News Corp. & $0.02\$$ &  & OMC & Omnicom Group Inc. & $0.03\$$\tabularnewline
    KLAC & KLA Corp. & $0.02\$$ &  & URBN & Urban Outfitters Inc. & $0.02\$$ &  & XRAY & DENTSPLY SIRONA Inc. & $0.03\$$\tabularnewline
    CIT & CIT Group Inc. & $0.02\$$ &  & RX & Recylex & $0.02\$$ &  & COP & ConocoPhillips & $0.03\$$\tabularnewline
    LEG & Leggett \& Platt Inc. & $0.02\$$ &  & BBY & Best Buy Inc. & $0.02\$$ &  & PCS & MetroPCS Communications & $0.03\$$\tabularnewline
    WFMI & Whole Foods Market Inc. & $0.02\$$ &  & CCL & Carnival Corp. & $0.02\$$ &  & RHI & Robert Half International Inc. & $0.03\$$\tabularnewline
    MS & Morgan Stanley & $0.02\$$ &  & WMI & WMI Investment Corp. & $0.02\$$ &  & SEE & Sealed Air Corp. & $0.03\$$\tabularnewline
    VRSN & VeriSign Inc. & $0.02\$$ &  & POM & Polymet Mining Corp. & $0.02\$$ &  & COST & Costco Wholesale Corp. & $0.03\$$\tabularnewline
    AN & AutoNation Inc. & $0.02\$$ &  & ERTS & Electronic Arts Inc. & $0.03\$$ &  & FIS & Fidelity National Info. Services & $0.03\$$\tabularnewline
    RHT & Red Hat Inc. & $0.02\$$ &  & MBI & MBIA Inc. & $0.03\$$ &  & PKI & PerkinElmer Inc. & $0.03\$$\tabularnewline
    CTXS & Citrix Systems Inc. & $0.02\$$ &  & ADM & Archer-Daniels-Midland Co. & $0.03\$$ &  & BMC & BMC Software Inc. & $0.03\$$\tabularnewline
    MDT & Medtronic plc. & $0.02\$$ &  & PEP & PepsiCo Inc. & $0.03\$$ &  & RRD & R.R. Donnelley \& Sons Co. & $0.03\$$\tabularnewline
    NBR & Nabors Industries & $0.02\$$ &  & CBG & CBRE Group Inc. & $0.03\$$ &  & UTX & United Technologies Corp. & $0.03\$$\tabularnewline
    MOLX & Molex Inc. & $0.02\$$ &  & IR & Ingersoll Rand Inc. & $0.03\$$ &  & D & Dominion Energy Inc. & $0.03\$$\tabularnewline
    GILD & Gilead Sciences Inc. & $0.02\$$ &  & HNZ & Heinz Company & $0.03\$$ &  & PBI & Pitney Bowes Inc. & $0.03\$$\tabularnewline
    CHK & Chesapeake Energy  & $0.02\$$ &  & CTX & Qwest Corp. & $0.03\$$ &  & ACAS & American Capital Ltd. & $0.03\$$\tabularnewline
    TJX & TJX Companies Inc. & $0.02\$$ &  & TSO & Tesoro Corp. & $0.03\$$ &  & K & Kellogg Company & $0.03\$$\tabularnewline
    AMGN & Amgen Inc. & $0.02\$$ &  & IGT & International Game Tech. & $0.03\$$ &  & JCP & J. C. Penney Company & $0.03\$$\tabularnewline
    SWY & Safeway Inc. & $0.02\$$ &  & WYN & Wynnstay Group PLC & $0.03\$$ &  & AMT & American Tower Corp. & $0.03\$$\tabularnewline
    XOM & Exxon Mobil Corp. & $0.02\$$ &  & GT & The Goodyear Tire \& Rubber & $0.03\$$ &  & ALL & Allstate Corp. & $0.03\$$\tabularnewline
    STZ & Constellation Brands & $0.02\$$ &  & JCI & Johnson Controls Int. & $0.03\$$ &  & MWV & MeadWestvaco Corp. & $0.03\$$\tabularnewline
    ADSK & Autodesk Inc. & $0.02\$$ &  & JWN & Nordstrom Inc. & $0.03\$$ &  & HRB & H\&R Block Inc. & $0.03\$$\tabularnewline
    LLY & Eli Lilly and Company & $0.02\$$ &  & FRX & Fennec Pharmaceutical Inc. & $0.03\$$ &  & NYT & New York times Company & $0.04\$$\tabularnewline
    CTAS & Cintas Corp. & $0.02\$$ &  & FHN & First Horizon National Corp. & $0.03\$$ &  & RDC & Redcape Hotel Group & $0.04\$$\tabularnewline
    LIZ & Liz Claiborne Inc. & $0.02\$$ &  & ABT & Abbott Laboratories & $0.03\$$ &  & PTV & Pactiv Company & $0.04\$$\tabularnewline
    GCI & Gannett Co. Inc. & $0.02\$$ &  & ADP & Automatic Data Processing & $0.03\$$ &  & FISV & Fiserv Inc. & $0.04\$$\tabularnewline
    AXP & American Express Co. & $0.02\$$ &  & PBG & Pacific Brands Ltd. & $0.03\$$ &  & EXPD & Expeditors Int.of Washington & $0.04\$$\tabularnewline
    TIE & Titanium Metals Corp. & $0.02\$$ &  & ROST & Ross Stores Inc. & $0.03\$$ &  & BBT & BB\&T Corp. & $0.04\$$\tabularnewline
    SAI & SAIC Inc. & $0.02\$$ &  & KBH & KB Home & $0.03\$$ &  & PCG & Pacific Gas \& Electric Co. & $0.04\$$\tabularnewline
    PDCO & Patterson Companies  & $0.02\$$ &  & YUM & Yum! Brands Inc. & $0.03\$$ &  & BIG & Big Lots Inc. & $0.04\$$\tabularnewline
    WYE & Wyeth Inc. & $0.02\$$ &  & MAR & Marriott International & $0.03\$$ &  & KMX & CarMax Inc. & $0.04\$$\tabularnewline
    COH & Cochlear Ltd. & $0.02\$$ &  & STJ & St Jude Medical Inc. & $0.03\$$ &  & TSS & Total System Services Inc. & $0.04\$$\tabularnewline
    CVG & Convergys Corp. & $0.02\$$ &  & FDO & Family Dollar Stores Inc. & $0.03\$$ &  & BK & The Bank of N. Y. Mellon Corp. & $0.04\$$\tabularnewline
    WDC & Western Digital Corp. & $0.02\$$ &  & ED & Consolidated Edison Inc. & $0.03\$$ &  & TEL & Tellurian Inc. & $0.04\$$\tabularnewline
    AVP & Avon Products Inc. & $0.02\$$ &  & UNM & Unum Group & $0.03\$$ &  & KSS & Kohl's Corp. & $0.04\$$\tabularnewline
    A & Agilent Technologies  & $0.02\$$ &  & ORLY & O'Reilly Automotive Inc. & $0.03\$$ &  & CAT & Caterpillar Inc. & $0.04\$$\tabularnewline
    JNY & Jones Apparel Group & $0.02\$$ &  & SVU & SUPERVALU Inc. & $0.03\$$ &  & HSY & The Hershey Company & $0.04\$$\tabularnewline
    SLM & SLM Corp. & $0.02\$$ &  & EMR & Emerson Electric Company & $0.03\$$ &  & GIS & General Mills Inc. & $0.04\$$\tabularnewline
    \end{tabular}

    \label{tab:spread_comp_2}
    \begin{tablenotes}
    \item[1] Average spread from 9:40:00 to 15:50:00 New York time during 2008.
    \end{tablenotes}

    \end{threeparttable}
    \end{table}
    \end{landscape}
    \clearpage
}

\afterpage{
    \clearpage
    \begin{landscape}
    \begin{table}
    \begin{threeparttable}
    \caption{Information of the stocks in Group 1 and 2.}

    \begin{tabular}{lllllllllll}
    \multicolumn{3}{c}{Group 1} &  & Symbol & Company & Spread\tnote{1} &  & Symbol & Company & Spread\tnote{1}\tabularnewline
    \cline{1-3} \cline{5-7} \cline{9-11}
    Symbol & Company & Spread\tnote{1} &  & DOV & Dover Corp. & $0.05\$$ &  & CAM & Corporate Actions Middleware & $0.06\$$\tabularnewline
    \cline{1-3}
    HAS & Hasbro Inc. & $0.04\$$ &  & VIAB & Viacom Inc. & $0.05\$$ &  & RAI & Reynolds American Inc. & $0.06\$$\tabularnewline
    XTO & XTO Energy Inc. & $0.04\$$ &  & ABC & AmerisourceBergen Corp. & $0.05\$$ &  & NOC & Northrop Grumman Corp. & $0.06\$$\tabularnewline
    PPL & PPL Corp. & $0.04\$$ &  & APC & Anadarko Petroleum Corp. & $0.05\$$ &  & PLL & Piedmont Lithium Ltd. & $0.06\$$\tabularnewline
    HOG & Harley-Davidson Inc. & $0.04\$$ &  & CBE & Cooper Industries & $0.05\$$ &  & SYK & Stryker Corp. & $0.06\$$\tabularnewline
    UPS & United Parcel Service & $0.04\$$ &  & FAST & Fastenal Company & $0.05\$$ &  & SRE & Sempra Energy & $0.06\$$\tabularnewline
    HSP & Hospira Inc. & $0.04\$$ &  & MHP & McGraw-Hill Companies Inc. & $0.05\$$ &  & TIF & Tiffany \& Co. & $0.06\$$\tabularnewline
    CTL & CenturyLink Inc. & $0.04\$$ &  & AMZN & Amazon.com Inc. & $0.05\$$ &  & NUE & Nucor Corp. & $0.06\$$\tabularnewline
    TDC & Teradata Corp. & $0.04\$$ &  & EQR & Equity Residential & $0.05\$$ &  & OXY & Occidental Petroleum Corp. & $0.06\$$\tabularnewline
    BA & Boeing Company & $0.04\$$ &  & CL & Colgate-Palmolive Company & $0.05\$$ &  & FPL & First Trust New Opportunities & $0.06\$$\tabularnewline
    BAX & Baxter International Inc. & $0.04\$$ &  & ECL & Ecolab Inc. & $0.05\$$ &  & ESRX & Express Scripts Holding Co. & $0.06\$$\tabularnewline
    PGN & Progress Energy Inc. & $0.04\$$ &  & MKC & McCormick \& Company Inc. & $0.05\$$ &  & AIV & Apartment Investment Co. & $0.06\$$\tabularnewline
    DRI & Darden Restaurants Inc. & $0.04\$$ &  & AOC & Aon Corp. & $0.05\$$ &  & BHI & Boulevard Holdings Inc. & $0.06\$$\tabularnewline
    JNS & Janus Capital Group Inc. & $0.04\$$ &  & TXT & Textron Inc. & $0.05\$$ &  & FCX & Freeport-McMoRan Inc. & $0.06\$$\tabularnewline
    BMS & Bristol-Myers Squibb & $0.04\$$ &  & CSX & CSX Corp. & $0.05\$$ &  & APOL & Apollo Group Inc. & $0.06\$$\tabularnewline
    RTN & Raytheon Company & $0.04\$$ &  & PNW & Pinnacle West Capital Corp. & $0.05\$$ &  & SCG & Scentre Group Ltd. & $0.06\$$\tabularnewline
    CVC & Cablevision Systems & $0.04\$$ &  & KIM & Kimco Realty Corp. & $0.05\$$ &  & CB & Chubb Limited & $0.06\$$\tabularnewline
    PWR & Quanta Services Inc. & $0.04\$$ &  & WPI & Watson Pharmaceuticals & $0.05\$$ &  & IBM & Int. Business Machines Corp. & $0.06\$$\tabularnewline
    LXK & Lexmark International & $0.04\$$ &  & MET & Metlife Inc. & $0.05\$$ &  & GME & GameStop Corp. & $0.06\$$\tabularnewline
    CELG & Celgene Corp. & $0.04\$$ &  & UST & ProShares Ultra & $0.05\$$ &  & DE & Deere \& Company & $0.06\$$\tabularnewline
    MMM & 3M Company & $0.04\$$ &  & TMO & Thermo Fisher Scientific Inc. & $0.05\$$ &  & LNC & Lincoln National Corp. & $0.06\$$\tabularnewline
    RSG & Republic Services Inc. & $0.04\$$ &  & HCP & HCP Inc. & $0.05\$$ &  & AVY & Avery Dennison Corp. & $0.07\$$\tabularnewline
    DNR & Denbury Resources Inc. & $0.04\$$ &  & SIAL & Sigma-Aldrich Corp. & $0.05\$$ &  & MCK & McKesson Corp. & $0.07\$$\tabularnewline
    MTW & Manitowoc Company & $0.04\$$ &  & FLIR & FLIR Systems Inc. & $0.05\$$ &  & PLD & Prologis Inc. & $0.07\$$\tabularnewline
    NE & Noble Corp. & $0.04\$$ &  & EL & Estee Lauder Companies & $0.05\$$ &  & DGX & Quest Diagnostics Inc. & $0.07\$$\tabularnewline
    NDAQ & Nasdaq Inc. & $0.04\$$ &  & AYE & Allegheny Energy Inc. & $0.05\$$ &  & CERN & Cerner Corp. & $0.07\$$\tabularnewline
    CINF & Cincinnati Financial & $0.04\$$ &  & CI & Cigna Corp. & $0.05\$$ &  & IRM & Irom Mountain Inc. & $0.07\$$\tabularnewline
    BIIB & Biogen Inc. & $0.04\$$ &  & NFLX & Netflix Inc. & $0.05\$$ &  & COL & Rockwell Collins Inc. & $0.07\$$\tabularnewline
    KMB & Kimberly-Clark Corp. & $0.04\$$ &  & CHRW & C.H. Robinson Worlwide Inc. & $0.05\$$ &  & NU & NeutriSci International Inc. & $0.07\$$\tabularnewline
    EIX & Edison International & $0.04\$$ &  & GENZ & Genzyme Corp. & $0.05\$$ &  & CVH & Coventry Health Care Inc. & $0.07\$$\tabularnewline
    NRG & NRG Energy Inc. & $0.04\$$ &  & DDR & DDR Corp. & $0.05\$$ &  & WFR & MEMC Electronic Materials & $0.07\$$\tabularnewline
    IVZ & Invesco Ltd. & $0.04\$$ &  & WFT & West Fraser Timber Co. Ltd. & $0.05\$$ &  & AGN & Allergan PLC. & $0.07\$$\tabularnewline
    AEE & Ameren Corp. & $0.04\$$ &  & MHS & Medco Health Solutions Inc. & $0.05\$$ &  & GAS & GAS & $0.07\$$\tabularnewline
    AAPL & Apple Inc. & $0.04\$$ &  & DTE & DTE Energy Company & $0.05\$$ &  & APH & Amphenol Corp. & $0.07\$$\tabularnewline
    CAH & Cardinal Health Inc. & $0.04\$$ &  & TRV & The Travelers Companies & $0.05\$$ &  & EXC & Exelon Corp. & $0.07\$$\tabularnewline
    MCO & Moody's Corp. & $0.04\$$ &  & NYX & NYSE Group IPO & $0.05\$$ &  & WEC & WEC Energy Group Inc. & $0.07\$$\tabularnewline
    \multicolumn{3}{c}{Group 2} &  & COF & Capital One Financial Corp. & $0.06\$$ &  & AFL & AFLAC Inc. & $0.07\$$\tabularnewline
    \cline{1-3}
    Symbol & Company & Spread\tnote{1} &  & WLP & WellPoint Inc. & $0.06\$$ &  & NOV & National Oilwell Varco Inc. & $0.07\$$\tabularnewline
    \cline{1-3}
    HOT & Hot Topic Inc. & $0.05\$$ &  & TWC & TWC Entreprises Ltd. & $0.06\$$ &  & SUN & Sunoco LP & $0.07\$$\tabularnewline
    GPC & Genuine Parts Co & $0.05\$$ &  & SWN & Southwestern Energy Co. & $0.06\$$ &  & BTU & Peabody Energy Corp. & $0.07\$$\tabularnewline
    PEG & Public Service Ent. & $0.05\$$ &  & NSC & Norfolk Southern Corp. & $0.06\$$ &  & ROK & Rockwell Automation Inc. & $0.07\$$\tabularnewline
    EFX & Equifax Inc. & $0.05\$$ &  & CMA & Comerica Inc. & $0.06\$$ &  & PCL & Plum Creek Timber Co. Inc. & $0.07\$$\tabularnewline
    COV & Covidien Ltd. & $0.05\$$ &  & NKE & Nike Inc. & $0.06\$$ &  & JOYG & Joy Global Inc. & $0.07\$$\tabularnewline
    AET & Aetna Inc. & $0.05\$$ &  & EQ & Equillium Inc. & $0.06\$$ &  & AMP & Ameriprise Financial Inc. & $0.07\$$\tabularnewline
    XL & XL Group Ltd. & $0.05\$$ &  & CLX & Clorox Company & $0.06\$$ &  & TAP & Molson Coors Beverage Co. & $0.07\$$\tabularnewline
    \end{tabular}

    \label{tab:spread_comp_3}
    \begin{tablenotes}
    \item[1] Average spread from 9:40:00 to 15:50:00 New York time during 2008.
    \end{tablenotes}

    \end{threeparttable}
    \end{table}
    \end{landscape}
    \clearpage
}

\afterpage{
    \clearpage
    \begin{landscape}
    \begin{table}
    \begin{threeparttable}
    \caption{Information of the stocks in Group 2 and 3.}

    \begin{tabular}{lllllllllll}
    \multicolumn{3}{c}{Group 2} &  & Symbol & Company & Spread\tnote{1} &  & Symbol & Company & Spread\tnote{1}\tabularnewline
    \cline{1-3} \cline{5-7} \cline{9-11}
    Symbol & Company & Spread\tnote{1} &  & ATI & Allegheny Technologies Inc. & $0.09\$$ &  & MUR & Murphy Oil Corp. & $0.13\$$\tabularnewline
    \cline{1-3}
    GD & General Dynamics Corp. & $0.07\$$ &  & ETN & Eaton Corp. & $0.09\$$ &  & ROP & Roper Technologies Inc. & $0.13\$$\tabularnewline
    CSC & Computer Sciences Corp. & $0.07\$$ &  & VTR & Ventas Inc. & $0.09\$$ &  & JEC & Jura Energy Corp. & $0.13\$$\tabularnewline
    FII & Federated Investors Inc. & $0.07\$$ &  & \multicolumn{3}{c}{Group 3} &  & HES & Hess Corp. & $0.13\$$\tabularnewline
    \cline{5-7}
    HIG & Hartford Fin. Services Group & $0.07\$$ &  & Symbol & Company & Spread\tnote{1} &  & ETR & Entergy Corp. & $0.13\$$\tabularnewline
    \cline{5-7}
    ITT & ITT Inc. & $0.07\$$ &  & HAR & Harman Int. Industries Inc. & $0.10\$$ &  & MON & Monsanto Company & $0.14\$$\tabularnewline
    TROW & T. Rowe Price Group Inc. & $0.07\$$ &  & NFX & Nasdaq Futures & $0.10\$$ &  & DV & Dolly Varden Silver Corp. & $0.14\$$\tabularnewline
    MDP & Meredith Corp. & $0.07\$$ &  & EQT & EQT Corp. & $0.10\$$ &  & VFC & V.F. Corp. & $0.14\$$\tabularnewline
    GR & Goodrich Corp. & $0.07\$$ &  & HRS & Harris Corp. & $0.10\$$ &  & GWW & W. W. Grainger Inc. & $0.14\$$\tabularnewline
    AKS & AK Steel Holding Corp. & $0.07\$$ &  & COG & Cabot Oil \& Gas Corp. & $0.10\$$ &  & WHR & Whirlpool Corp. & $0.15\$$\tabularnewline
    PFG & Principal Financial Group & $0.07\$$ &  & PPG & PPG Industries Inc. & $0.10\$$ &  & MOS & Mosaic Company & $0.15\$$\tabularnewline
    HUM & Humana Inc. & $0.07\$$ &  & MEE & Massey Energy Company & $0.10\$$ &  & SPG & Simon Property Group & $0.15\$$\tabularnewline
    STI & SunTrust Banks Inc. & $0.07\$$ &  & HP & Helmerich \& Payne Inc. & $0.10\$$ &  & EOG & EOG Resources Inc. & $0.15\$$\tabularnewline
    CMI & Cummins Inc. & $0.08\$$ &  & DVN & Devon Energy Corp. & $0.10\$$ &  & VMC & Vulcan Materials Co & $0.16\$$\tabularnewline
    STR & Questar Corp. & $0.08\$$ &  & IFF & Int. Flavors \& Fragrances & $0.10\$$ &  & X & United States Steel Corp. & $0.16\$$\tabularnewline
    ZMH & Zimmer Holdings Inc. & $0.08\$$ &  & LH & Laboratory Corp. & $0.10\$$ &  & SHLD & Sears Holding & $0.16\$$\tabularnewline
    FO & Fortune Brands Inc. & $0.08\$$ &  & UNP & Union Pacific Corp. & $0.10\$$ &  & LLL & L3 Technologies Inc. & $0.17\$$\tabularnewline
    SII & Sprott Inc. & $0.08\$$ &  & EW & Edwards Lifesciences Corp. & $0.10\$$ &  & WYNN & Wynn Resorts Ltd. & $0.17\$$\tabularnewline
    HRL & Hormel Foods Corp. & $0.08\$$ &  & STT & State Street Corp. & $0.10\$$ &  & BXP & Boston Properties Inc. & $0.17\$$\tabularnewline
    OI & O-I Glass Inc. & $0.08\$$ &  & RRC & Range Resources Corp. & $0.10\$$ &  & PSA & Public Storaga & $0.17\$$\tabularnewline
    HCN & Welltower Inc. & $0.08\$$ &  & SJM & J. M. Smucker Company & $0.10\$$ &  & PCP & Precision Castparts Corp. & $0.18\$$\tabularnewline
    ANF & Abercrombie \& Fitch Co. & $0.08\$$ &  & SNA & Snap-On Inc. & $0.11\$$ &  & VNO & Vornado Realty & $0.18\$$\tabularnewline
    LM & Legg Mason Inc. & $0.08\$$ &  & PNC & PNC Fin. Services Group & $0.11\$$ &  & DNB & Dun \& Bradstreet Corp. & $0.19\$$\tabularnewline
    FDX & FedEx Corp. & $0.08\$$ &  & CEG & Constellation Energy Group & $0.11\$$ &  & CLF & Cleveland-Cliffs Inc. & $0.19\$$\tabularnewline
    PRU & Prudential Financial Inc. & $0.08\$$ &  & LMT & Lockheed Martin Corp. & $0.11\$$ &  & FLR & Fluor Corp. & $0.20\$$\tabularnewline
    DHR & Danaher Corp. & $0.08\$$ &  & FTI & TechnipFMC PLC & $0.11\$$ &  & PCLN & Priceline Group Inc. & $0.20\$$\tabularnewline
    WY & Weyerhaeuser Company & $0.08\$$ &  & CNX & CNX Resources Corp. & $0.11\$$ &  & MTB & M\&T Banc Corp. & $0.20\$$\tabularnewline
    OKE & ONEOK Inc. & $0.08\$$ &  & ARG & Amerigo Resources Ltd. & $0.11\$$ &  & AVB & AvalonBay Comm. Inc. & $0.21\$$\tabularnewline
    SRCL & Stericycle Inc. & $0.09\$$ &  & DVA & DaVita Inc. & $0.11\$$ &  & BEN & Franklin Resources Inc. & $0.21\$$\tabularnewline
    LUK & Leucadia National Corp. & $0.09\$$ &  & GS & Goldman Sachs Group Inc. & $0.11\$$ &  & DO & Diamond Offshore Drilling & $0.23\$$\tabularnewline
    BLL & Ball Corp. & $0.09\$$ &  & TMK & Torchmark Corp. & $0.12\$$ &  & CF & CF Industries Holdings & $0.24\$$\tabularnewline
    FE & FirstEnergy Corp. & $0.09\$$ &  & EMN & Eastman Chemical Co. & $0.12\$$ &  & AZO & AutoZone Inc. & $0.25\$$\tabularnewline
    SWK & Stanley Black \& Decker Inc. & $0.09\$$ &  & RIG & Transocean Ltd. & $0.12\$$ &  & FLS & Flowserve Corp. & $0.27\$$\tabularnewline
    ESV & Ensco PLC & $0.09\$$ &  & PX & Pelangio Exploration Inc. & $0.12\$$ &  & ICE & Intercontinental Exc. Inc. & $0.28\$$\tabularnewline
    SHW & Sherwin-Williams Company & $0.09\$$ &  & RL & Ralph Lauren Corp. & $0.12\$$ &  & MA & Mastercard Inc. & $0.38\$$\tabularnewline
    BDX & Becton, Dickinson and Co. & $0.09\$$ &  & R & Ryder System & $0.12\$$ &  & FSLR & First Solar Inc. & $0.38\$$\tabularnewline
    VAR & Varian Medical Systems Inc. & $0.09\$$ &  & MIL & Millipore Corp. & $0.12\$$ &  & CMG & Chipotle Mexican Grill & $0.38\$$\tabularnewline
    TEG & Ten Ent. Group PLC & $0.09\$$ &  & CRM & Salesforce.com Inc. & $0.12\$$ &  &  &  & \tabularnewline
    PH & Parker-Hannifin Corp. & $0.09\$$ &  & NTRS & Northern Trust Corp. & $0.12\$$ &  &  &  & \tabularnewline
    ACE & ACE Ltd. & $0.09\$$ &  & APD & Air Products and Chemicals & $0.12\$$ &  &  &  & \tabularnewline
    BNI & Burlington Northern Santa Fe & $0.09\$$ &  & ANR & Alpha Natural Resources & $0.13\$$ &  &  &  & \tabularnewline
    ACS & Affiliated Computer Services & $0.09\$$ &  & NBL & Noble Energy Inc. & $0.13\$$ &  &  &  & \tabularnewline
    PXD & Pioneer Natural Resources  & $0.09\$$ &  & APA & Apache Corp. & $0.13\$$ &  &  &  & \tabularnewline
    AIZ & Assurant Inc. & $0.09\$$ &  & FMC & FMC Corp. & $0.13\$$ &  &  &  & \tabularnewline
    WAT & Waters Corp. & $0.09\$$ &  & BDK  & Black and Decker Corp. & $0.13\$$ &  &  &  & \tabularnewline
    \end{tabular}

    \label{tab:spread_comp_4}
    \begin{tablenotes}
    \item[1] Average spread from 9:40:00 to 15:50:00 New York time during 2008.
    \end{tablenotes}

    \end{threeparttable}
    \end{table}
    \end{landscape}
    \clearpage
}